\documentclass[11pt]{article}  %envcountsame
\usepackage{amsfonts,graphicx}
\newtheorem{theorem}{T{\sc HEOREM}}
\newtheorem{lemma}{L{\sc EMMA}}
\newtheorem{proposition}{P{\sc ROPOSITION}}
\newtheorem{corollary}{C{\sc OROLLARY}}

\def\bp{\noindent{\it Proof. }}
\def\ep{\noindent{\hfill \fbox{}}}
\def\remark{\noindent{\bf Remark. }}
\def\definition{\noindent{\bf Definition. }}

\def\pic{\rm Pic}
\newcommand{\ol}[1]{\overline{#1}}
\newcommand{\mapright}[1]{%
   \smash{\mathop{%
   \hbox to 1cm{\rightarrowfill}}\limits^{#1}}}
\newcommand{\mapleft}[1]{%
   \smash{\mathop{%
   \hbox to 1cm{\leftarrowfill}}\limits^{#1}}}
\newcommand{\maplleft}[2]{%
   \smash{\mathop{%
   \hbox to 1cm{\leftarrowfill}}\limits_{#1}^{#2}}}

\begin{document}

\title{Discrete dynamical systems associated with 
root systems of indefinite type}
\date{}
\maketitle
\begin{center}
{Graduate School of Mathematical Sciences, University of
Tokyo, Komaba 3-8-1, Meguro-ku, Tokyo 153-8914, Japan}\\
\end{center}

\begin{abstract}
A geometric charactrization of the equation found by Hietarinta and 
Viallet, which satisfies the singularity confinement criterion but 
which exhibits chaotic behavior, is presented. It is shown that this equation 
can be lifted to an 
automorphism of a certain rational surface and 
can therefore be considered to be a realization of a Cremona isometry
on the Picard group of the surface.
It is also shown that the group of Cremona isometries is isomorphic
to an extended Weyl group of indefinite type.
A method to construct 
the mappings associated with some root systems of indefinite type 
is also presented. 
\end{abstract}

\section{Introduction}

The singularity confinement method has been proposed
by Grammaticos et al. \cite{grp} as a
criterion for the integrability of (finite or infinite dimensional) 
discrete dynamical
systems.  The singularity confinement method demands that
when singularities appear due to particular initial values
such singularities should disappear after a finite number of iteration steps,
in which case the information on the initial values ought to be recovered 
(hence the dynamical system
has to be invertible). 

However ``counter examples'' were found by Hietarinta and Viallet
\cite{hv}. These mappings satisfy the singularity confinement criterion
but the orbits of their solutions exhibit chaotic behavior.
The authors of \cite{hv} introduced the notion of algebraic entropy in order to test the degree of complexity of successive iterations.
The algebraic entropy is defined as $s:=\lim_{n\to \infty}\log(d_n)/n$
where $d_n$ is the degree of the $n$th iterate.
This notion is linked to Arnold's complexity since
the degree of a mapping gives the intersection number of the image
of a line and a hyperplane. While the degree grows exponentially for
a generic mapping, it was shown that it only grows
polynomially  for a large class of integrable mappings
\cite{arnord,bv,hv,otgr}.

The discrete Painlev\'{e} equations have been extensively studied \cite{js,rgh}.
Recently it was shown by Sakai \cite{sakai} that all
of (from the point of view of symmetries) these are
obtained by studying rational surfaces in connection with the extended
affine Weyl groups. 

Surfaces obtained by successive blow-ups \cite{hartshorne}
of ${\mathbb P}^2$ or
${\mathbb P}^1 \times {\mathbb P}^1$ have been studied by several authors 
by means of connections 
between the Weyl groups and the groups of Cremona isometries on the 
Picard group of the surfaces \cite{cd,dolgachev,dolgachev2}.
Here, the Picard group of a rational surface $X$ is the group 
of isomorphism classes
of invertible sheaves on $X$ and
it is isomorphic to the group of linear equivalent classes of divisors on $X$.
A Cremona isometry is an isomorphism of the Picard group such that
a) it preserves the intersection number of any pair of divisors, 
b) it preserves the canonical divisor $K_X$ and 
c) it leaves the set of effective classes of divisors invariant.
In the case where $9$ points (in the case of ${\mathbb P}^2$, $8$ points
in the case of ${\mathbb P}^1 \times {\mathbb P}^1$) are blown up, 
if the points are in general position the group of Cremona 
isometries becomes isomorphic with 
an extension of the Weyl group of type $E_8^{(1)}$. 
In case the $9$ points are not in general 
position,
the classification of  connections between the group of Cremona isometries
and the extended affine Weyl groups
was first studied by Looijenga \cite{looijenga} and
more generally by Sakai.
Birational (bi-meromorphic) mappings 
on ${\mathbb P}^2$ (or
${\mathbb P}^1 \times {\mathbb P}^1$) are obtained by interchanging 
the procedure of blow downs.  Discrete
Painlev\'{e} equations are recovered as the birational mappings
corresponding to the translations
of affine Weyl groups. It was shown that for every Painlev\'e equation
in Sakai's list the order of the $n$th iterate is at most $O(n^2)$ 
\cite{takenawa2}.

Our aim in this paper is to characterize some birational mappings  
which satisfy the singularity confinement criterion but exhibit chaotic
behavior from the point of view of the theory of rational surfaces. 
Considering one such mapping and the space of its initial values,  
we obtain a rational surface associated with some root system of 
hyperbolic type. 
Conversely, we recover
the mapping from the surface and consequently obtain the extension of 
the mapping to its non-autonomous version. 
It is important to remark that this method also allows the construction
of other mappings starting from suitable rational surfaces.
We also show some other examples of such constructions. 

In Section 2, we start from one of the mappings found by
Hietarinta and Viallet (we call it the HV eq. in this paper) 
and construct the space such that
the mappings is lifted to an automorphism, i.e. bi-holomorphic mapping, 
of the surface. The mapping $\varphi'$ is called a mapping lifted 
from the mapping $\varphi$ if $\varphi'$ coincides with $\varphi$
on any point where $\varphi$ is defined. 
For this purpose we compactify 
the original space of initial values, ${\mathbb C}^2$, to 
${\mathbb P}^1 \times {\mathbb P}^1$
and blow-up $14$ times.

In Section 3, we study the symmetry of the space of initial values.
We show that the group of all the Cremona isometries of the Picard group of the surface is isomorphic to 
an extended Weyl group of hyperbolic type.
As a corollary, we prove that there does not exist
any Cremona isometry whose action on the Picard group 
commutes with the action of the HV eq. except the action itself.  

In Section 4, we show a method to recover the HV eq. from
the surface as an element of the extended Weyl group. 
Each element of the extended Weyl group which acts on
the Picard group as a Cremona isometry, is realized as a
Cremona transformation (i.e. a birational mapping) on 
${\mathbb P}^1 \times {\mathbb P}^1$ by interchanging
the blow down structure. Here, a blow down structure is the sequence 
designating the procedure of blow downs.
As a result of this we obtain the non-autonomous version of the equation.

In Section 5, we discuss the construction of other mappings from
certain rational surfaces and show some examples which are associated
with root systems of indefinite type.

%%%%%%%%%%%%%%%%%%%%%%%%

\section{Construction of the space of initial values by blow-ups}

We consider the dynamical system written by the birational (bi-meromorphic)
mapping
\begin{eqnarray}\label{hv}
\varphi:&{\mathbb C}^2 & \to ~ {\mathbb C}^2 \nonumber \\
&\left(\begin{array}{c}
x_n\\y_n \end{array}\right)
& \mapsto ~
\left(\begin{array}{c}
x_{n+1}\\y_{n+1} \end{array}\right)
= \left(\begin{array}{c}
y_n\\ -x_n+y_n+a/y_n^2 \end{array}\right) \label{hv} 
\end{eqnarray}
where $a \in {\mathbb C}$ is a nonzero constant.
This equation was found by Hietarinta and Viallet \cite{hv} and we call
it the HV eq.. To test the singularity confinement,
let us assume $x_0\neq0$ and $y_0=\epsilon$ where $|\epsilon|\ll 1$.
With these initial values  we obtain the sequence:
\begin{eqnarray*}
\begin{array}{lll}
x_0&=&x_0\\
x_1=y_0&=&\epsilon\\
x_2=y_1&=&a\epsilon^{-2}-x_0+\epsilon\\
x_3=y_2&=&a\epsilon^{-2} -x_0+a^{-1}\epsilon^4 + O(\epsilon^6)\\
x_4=y_3&=&-\epsilon +2 a^{-1} \epsilon^{4} +4 x_0 a^{-2} \epsilon^{6}
              + O(\epsilon^7)\\
x_5=y_4&=&x_0 + 3 \epsilon + O(\epsilon^2)\\
x_6=y_5&=&(a x_0^{-2}+x_0) + O(\epsilon)\\
&\vdots&\quad.
\end{array}
\end{eqnarray*}

In this sequence singularities appear at $n=1$ as $\epsilon \to 0$ 
and disappear at $n=4$ and the information on the 
initial values is hidden in the coefficients of higher degree $\epsilon$.
However, taking suitable
rational functions of $x_n$ and $y_n$  we can find the information of the initial values as finite
values. The fact that the leading orders of
$(x_1^2 y_1-a)y_1,~ (x_2^3(y_2/x_2-1)^2-a)x_2$ and
$(x_3 y_3^2-a)x_3 $ become
$ - a x_0, -a x_0$ and $ - a x_0$ actually suggests that the HV eq. can be
lifted to an automorphism of a suitable rational surface  
(although these rational functions are of course
not uniquely determined). 

Let us consider the HV eq. $\varphi$ to be a mapping from 
the complex projective space 
${\mathbb P}^1({\mathbb C}) \times {\mathbb P}^1({\mathbb C})$
(= ${\mathbb P}^1 \times {\mathbb P}^1$) to itself.
We use the terminology {\it space of initial values} 
as follows (analogous to the space of initial values of Painlev\'{e} 
equations introduced by Okamoto\cite{okamoto}). \\

\definition  
A sequence of algebraic varieties $X_i$ 
is (or $X_i$ themselves are) called the space of initial values
for the sequence of rational mappings $\varphi_i$, 
if each $\varphi_i$ is lifted to an isomorphism from $X_i$ to $X_{i+1}$, 
for all $i$.\\

Our aim in this section is to construct the surface $X$ by
blow-ups ${\mathbb P}^1 \times {\mathbb P}^1$ 
such that $\varphi$ is lifted to an automorphism of $X$,
where the mapping $\varphi'$ is called a mapping lifted 
from the mapping $\varphi$ if $\varphi'$ coinsides with $\varphi$
on any point where $\varphi$ is defined.

\subsection{Regular mapping from $Y_1$ to 
${\mathbb P}^1 \times {\mathbb P}^1$}

Let the coordinates of ${\mathbb P}^1 \times {\mathbb P}^1$ be
$(x,y),(x,1/y),(1/x,y)$
and $(1/x,1/y)$ and let $x=\infty$ denote $1/x=0$.
We denote the HV eq. as
\begin{eqnarray}\label{ihv}
\varphi&:&(x,y)\mapsto
(\overline{x},\overline{y})=
(y, -x+y+a/y^2)
\end{eqnarray}
where $(\overline{x},\overline{y})$ means the image of $(x,y)$ by the mapping.
This mapping has two indeterminate points:
$(x,y)= (\infty,0),~(\infty,\infty)$.
By blowing up at these points we can ease the indeterminacy.

We denote blowing up at $(x,y)=(x_0,y_0)\in {\mathbb C}^2$:
\begin{eqnarray*}
&\{(x,y):x,y \in {\mathbb C}\}\\
\mapleft{\mu_{(x_0,y_0)}} &
\{(x-x_0,~y-y_0;~\zeta_1:\zeta_2)~|~x,y,\zeta_1,\zeta_2 \in {\mathbb C}
\end{eqnarray*}
by
\begin{eqnarray}\label{coblow}
 (x,y) \leftarrow (x-x_0,(y-y_0)/(x-x_0))\cup
((x-x_0)/(y-y_0), y-y_0). 
\end{eqnarray}
In this way, blowing up at $(x,y)=(x_0,y_0)$ gives meaning 
to $(x-x_0)/(y-y_0)$ at this point.

First we  blow-up at $(x,y)= (\infty,0)$,
$(1/x,y) \leftarrow (1/x,xy)\cup (1/xy,y)$
and denote the obtained surface by $Y_0$. Then
$\varphi$ is lifted to a rational mapping 
from $Y_0$ to ${\mathbb P}^1 \times {\mathbb P}^1$.
For example, in the new coordinates $\varphi$ is expressed as
\begin{eqnarray*} 
(u_1,v_1):=(1/x,xy) &\mapsto&
(\overline{x},\overline{y})=
(u_1 v_1, (-u_1 v_1^2  + u_1^3 v_1^3  + a)/(u_1^2 v_1^2))\\
(u_2,v_2):=(1/xy,y) &\mapsto&
(\overline{x},\overline{y})=(v_2, (-v_2 + u_2 v_2^3 +a u_2)/(u_2 v_2^2))
\end{eqnarray*}
This maps the exceptional
curve at $(x,y)=(\infty,0)$ , i.e.$u_1=0$ and $v_2=0$,  almost to 
$(\overline{x},\overline{y})=(\infty,0)$ but
has an indeterminate point on the exceptional curve:
$(u_2,v_2)=(0,0).$ Hence we have to blow-up again at this point.
In general it is known that, if there is a rational mapping $X \to X'$ where
$X$ and $X'$ are smooth projective algebraic varieties,
the procedure of blowing up can be completed in a finite number of steps, 
after 
which one obtains a smooth projective algebraic variety $Y$
such that the rational mapping is lifted to
a regular mapping from $Y$ to $X'$
(theorem of the elimination of indeterminacy \cite{hartshorne}).

Here we obtain the surface $Y_1$ defined by the following sequence 
of blow-ups. (For simplicity 
we take only one coordinate of (\ref{coblow}).) 
\begin{eqnarray*}
(x,y) &\maplleft{\mu_1}{(\infty,0)}& \left(\frac{1}{xy},y\right)
\maplleft{\mu_2}{(0,0)} \left(\frac{1}{xy},xy^2\right) \\
&\maplleft{\mu_3}{(0,a)}& \left(\frac{1}{xy}, xy(xy^2-a)\right)
\maplleft{\mu_4}{(0,0)} \left(\frac{1}{xy},x^2y^2(xy^2-a)\right)\\
(x,y) &\maplleft{\mu_9}{(\infty,\infty)}& \left(\frac{1}{x},\frac{x}{y}\right)
\maplleft{\mu_{10}}{(0,1)} \left(\frac{1}{x},x(\frac{x}{y}-1)\right)
\end{eqnarray*}
where $\mu_i$ denotes the $i$th blow-up.
Of course the above sequence is not unique since there is 
freedom in choosing the coordinates.

\subsection{Automorphism of $X$}
%%%%%%%%%%%

We have obtained a mapping from $Y_1$ to
${\mathbb P}^1 \times {\mathbb P}^1$ which is lifted from $\varphi$.
But our aim is to construct a rational surface $X$ such that
$\varphi$ is lifted to an automorphism of $X$.

First we construct the rational surface $Y_2$ such that
$\varphi$ is lifted to a regular mapping from $Y_2$ to $Y_1.$
For this purpose it is sufficient to eliminate the indeterminacy
of mapping from $Y_1$ to $Y_1$.
Consequently we obtain $Y_2$ defined by the following sequence of blow-ups.
\begin{eqnarray*}
\left(\frac{1}{x},x(\frac{x}{y}-1)\right)
&\maplleft{\mu_{11}}{(0,0)}& \left(\frac{1}{x^2(x/y-1)},  x(\frac{x}{y}-1)\right)\\
&\maplleft{\mu_{12}}{(0,0)}& 
\left(\frac{1}{x^2(x/y-1)},  x^3(\frac{x}{y}-1)^2\right)\\
&\maplleft{\mu_{13}}{(0,a)}& \left(\frac{1}{x^2(x/y-1)},
 x^2(\frac{x}{y}-1)(x^3(\frac{x}{y}-1)^2-a)\right)\\
&\maplleft{\mu_{14}}{(0,0)}&
\left(\frac{1}{x^2(x/y-1)},x^4(\frac{x}{y}-1)^2(x^3(\frac{x}{y}-1)^2-a)\right)
\end{eqnarray*}

Next eliminating the indeterminacy of mapping from $Y_2$ to $Y_2$,
we obtain $Y_3$ defined by the following sequence of blow-ups.
\begin{eqnarray*}
(x,y) &\maplleft{\mu_5}{\mbox{at}~(0,\infty)}& \left(x,\frac{1}{xy}\right)
\maplleft{\mu_6}{(0,0)} \left(x^2 y,\frac{1}{xy}\right) \\
&\maplleft{\mu_7}{(a,0)}& \left(xy(x^2 y-a),\frac{1}{xy}\right)
\maplleft{\mu_8}{(0,0)} \left(x^2y^2(x^2 y-a),\frac{1}{xy}\right)
\end{eqnarray*}

It can be shown that the mapping from $Y_3$ to $Y_3$ which is 
lifted from $\varphi$
does not have any indeterminate points.

To show this, let us define the total and proper transforms.\\

\definition
Let $S$ be the set of zero points of $\bigwedge_{i\in I} f_i(u,v)=0$,
where $(u,v)\in {\mathbb C}^2$ and the $\{f_i\}_{i\in I}$ is a finite 
set of polynomials,
and 
let $U_1:(u_1,v_1)$ and $U_2:(u_2,v_2)$ the new coordinates of blow-up at the
point $(u,v)=(a,b)$, i.e. $(u_1,v_1)=(u-a,(v-b)/(u-a)),$ $(u_2,v_2)=
((u-a)/(v-b),v-b)$. The {\it total transform} of $S$ is 
$$\{(u_1,v_1)\in U_1; f_i(u_1+a,u_1v_1+b)=0\}\cup
\{(u_2,v_2)\in U_2; f_i(u_2 v_2+a,v_2+b)=0\}$$
and the {\it proper transform} of $S$ is 
{\small
$$\{(u_1,v_1)\in U_1; \frac{ f_i(u_1+a,u_1v_1+b)}{u_1^m}=0\}\cup
\{(u_2,v_2)\in U_2; \frac{ f_i(u_2 v_2+a,v_2+b)}{v_2^n}=0\}$$}
where $m$ or $n$ is the maximum integer simplifying the respective equations for $u_1$ or $v_2$ 
respectively. \\

\noindent For example, by blowing up at $(u,v)=(0,0)$, the total transform of $u=0$
is $\{(u_1,v_1)\in U_1; u_1=0\}\cup\{(u_2,v_2)\in U_2;u_2 v_2=0\}$ 
and its proper transform is
$\{(u_1,v_1)\in U_1; 1=0\}(=\phi)\cup\{(u_2,v_2)\in U_2; u_2=0\}$.

\begin{figure}[ht]
\begin{center}
\includegraphics*[]{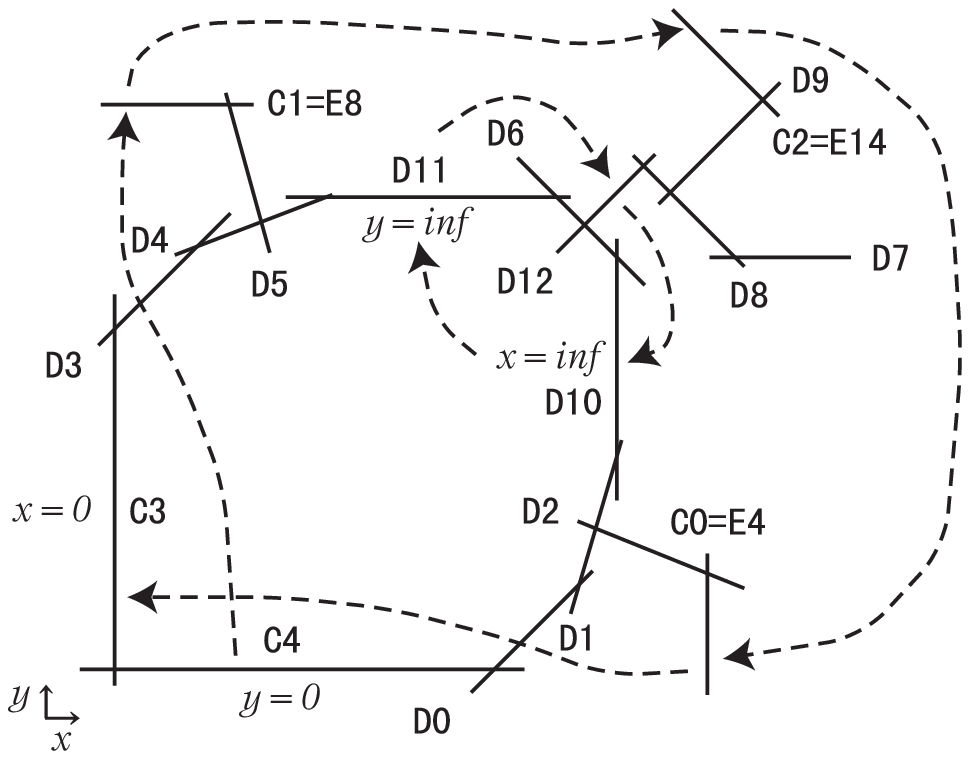}
\caption[]{}\label{proper}
\end{center}
\end{figure}

We denote the total transform of the point 
of the $i$th blow-up by $E_i$ and denote the proper transform of
the exceptional curves of the $i$th blow-up by 
\begin{eqnarray}\label{pexce}
D_0,D_1,D_2,C_0=E_4,D_3,D_4,D_5,C_1=E_8,D_6,D_{12},D_7,D_8,D_9,C_2=E_{14}.
\end{eqnarray}
Moreover we denote the proper transforms of $x=0$, $x=\infty$, $y=0$, 
$y=\infty$ as
\begin{eqnarray}\label{pline}
C_3,D_{10},C_4,D_{11}
\end{eqnarray}
(See Fig~\ref{proper}.)

The proper transforms of
$C_4,D_0,D_1,D_2$ and $C_0$ are written as

\begin{eqnarray*}
C_4:&(x,y)=(x,0)\\
D_0:&(u_1,v_1)=(u_1,0)\\
D_1:&(u_2,v_2)=(0,v_2)\\
D_2:&(u_3,v_3)=(0,v_3)\\
C_0:&(u_4,v_4)=(0,v_4)
\end{eqnarray*}
where $u_i$ and $v_i$ are the new coordinate of the $i$th blow-up
(more precisely, these express the total transforms of curves 
and we have to write each curve by using the coordinates of the last blow-up
but this makes the notation rather complicated)
and therefore the relations 

\begin{eqnarray*}
\begin{array}{ll}
x=1/(u_1 v_1),& y=v_1,\\
u_1=u_2 ,& v_1=u_2 v_2,\\
u_2=u_3 ,& v_2=u_3 v_3+a,\\
u_3=u_4,& v_3=u_4 v_4
\end{array}
\end{eqnarray*}
hold.  The proper transforms of $C_1, D_5,D_4,D_3$ and $C_3$ are
written as
\begin{eqnarray*}
C_1:&(u_8,v_8)=(u_8,0)\\
D_5:&(u_7,v_7)=(u_7,0)\\
D_4:&(u_6,v_6)=(u_6,0)\\
D_3:&(u_5,v_5)=(0,v_5)\\
C_3:&(x,1/y)=(0,1/y)
\end{eqnarray*}
and the relations 

\begin{eqnarray*}
\begin{array}{ll}
u_8=u_7/v_7,& v_8=v_7,\\
u_7=(u_6-a)/v_6,& v_7= v_6,\\
u_6=u_5/v_5,& v_6= v_5,\\
u_5=x,& v_5= 1/(x y) 
\end{array}
\end{eqnarray*}
hold. 

Using these relations one can calculate the images of the curves.
For example, in the case of $C_4$:
From the above relations and (\ref{ihv}) we can calculate 
$(\overline{u_8},\overline{v_8})$ using initial values corresponding to $C_1$ as
\begin{eqnarray*}
\left.(\overline{u_8},\overline{v_8})\right|_{(x,y)=(x,0)}&=&
\left. \left((-x + y)(a + y^2 (-x + y))^2,\frac{y}{a - x y^2 + y^3}\right)
\right|_{(x,y)=(x,0)} \\
&=&(-a^2 x,0)
\end{eqnarray*}
This then implies that the image of $C_4$ $( =\overline{C_4})$ is $C_1$.

Analogously, from the equation
\begin{eqnarray*}
&&\left.(\overline{u_7},\overline{v_7}) \right|_{(u_1,v_1)=(u_1,0)}\\
&=&\left. 
\left(\frac{(-1 + u_1 v_1^2)(a u_1 - v_1 + u_1 v_1^3)}{u_1^2},
\frac{u_1 v_1}{a u_1 - v_1 + u_1 v_1^3}\right) \right|_{(u_1,v_1)=(u_1,0)}\\ 
&=&\left(-\frac{a}{u_1},0\right)
\end{eqnarray*}
which implies $\overline{D_0}=D_5$. In this way we can show
that 
{\small
\begin{eqnarray}\label{actd}
\begin{array}{ll}
&(D_0,~D_1,~D_2,~D_3,~D_4,~D_5,~D_6,~D_7,~D_8,~
D_9,~D_{10},~D_{11},~D_{12},~C_0,~C_1,~C_2,~C_4)\\
\to&
(D_5,~D_4,~D_3,~D_7,~D_8,~
D_9,~D_6,~D_0,~D_1,~ D_2,~
D_{11},~D_{12},~D_{10},~C_3,~C_2,~C_0,~C_1).
\end{array}
\end{eqnarray}}

It is obvious that this mapping has an inverse 
(the mapping lifted from $\varphi^{-1}$). 
Hence we obtain the
following theorem.
\begin{theorem}
The HV eq. (\ref{hv}) can be lifted to an automorphism of $X (=Y_3)$.
\end{theorem}

%%%%%%%%%%%%%%%

\section{The Picard group and symmetry}

\subsection{Action on the Picard group}
We denote the (linear equivalent classes of) total transform of
$x={\rm constant},$ (or $y={\rm constant}$) on $X$ by $H_0$ (or $H_1$ respectively)
and the (linear equivalent classes of) total transform of the point 
of the $i$th blow-up by $E_i$. 
From \cite{hartshorne} we know that the Picard group of $X$, Pic($X$),
is
$${\rm Pic}(X)= {\mathbb Z}H_0 + {\mathbb Z}H_1+  {\mathbb Z}E_1+\cdots+{\mathbb Z}E_{14}$$
and that the canonical divisor of $X$, $K_X$, is
$$K_X=-2H_0 -2H_1 + E_1+\cdots+E_{14}$$
It is also known that the intersection form, i.e. the intersection numbers
of pairs of base elements, is
\begin{eqnarray}\label{isn}
H_i \cdot H_j = 1-\delta_{i,j},~ E_k\cdot E_l= -\delta_{k,l},~
H_i \cdot E_k=0~
\end{eqnarray}
where $\delta_{i,j}$ is $1$ if $i=j$ and $0$ if $i\neq j,$ and
the intersection numbers of any pairs of divisors are given by
their linear combinations. \\

\remark
Let $X$ be a rational surface.
It is known that Pic($X$), the group of isomorphism classes of 
invertible sheaves of $X$, is isomorphic to the following groups. \\
i)The group of linear equivalent classes of divisors on $X$.\\
ii)The group of numerically equivalent classes of divisors on $X$,
where divisors $D$ and $D'$ on $X$ are numerically equivalent
if and only if for any divisors $D''$ on $X$, $D\cdot D''=D'\cdot D''$
holds. \\
Hence we identify them in this paper. \\

\begin{figure}[ht]
\begin{center}
\includegraphics*[]{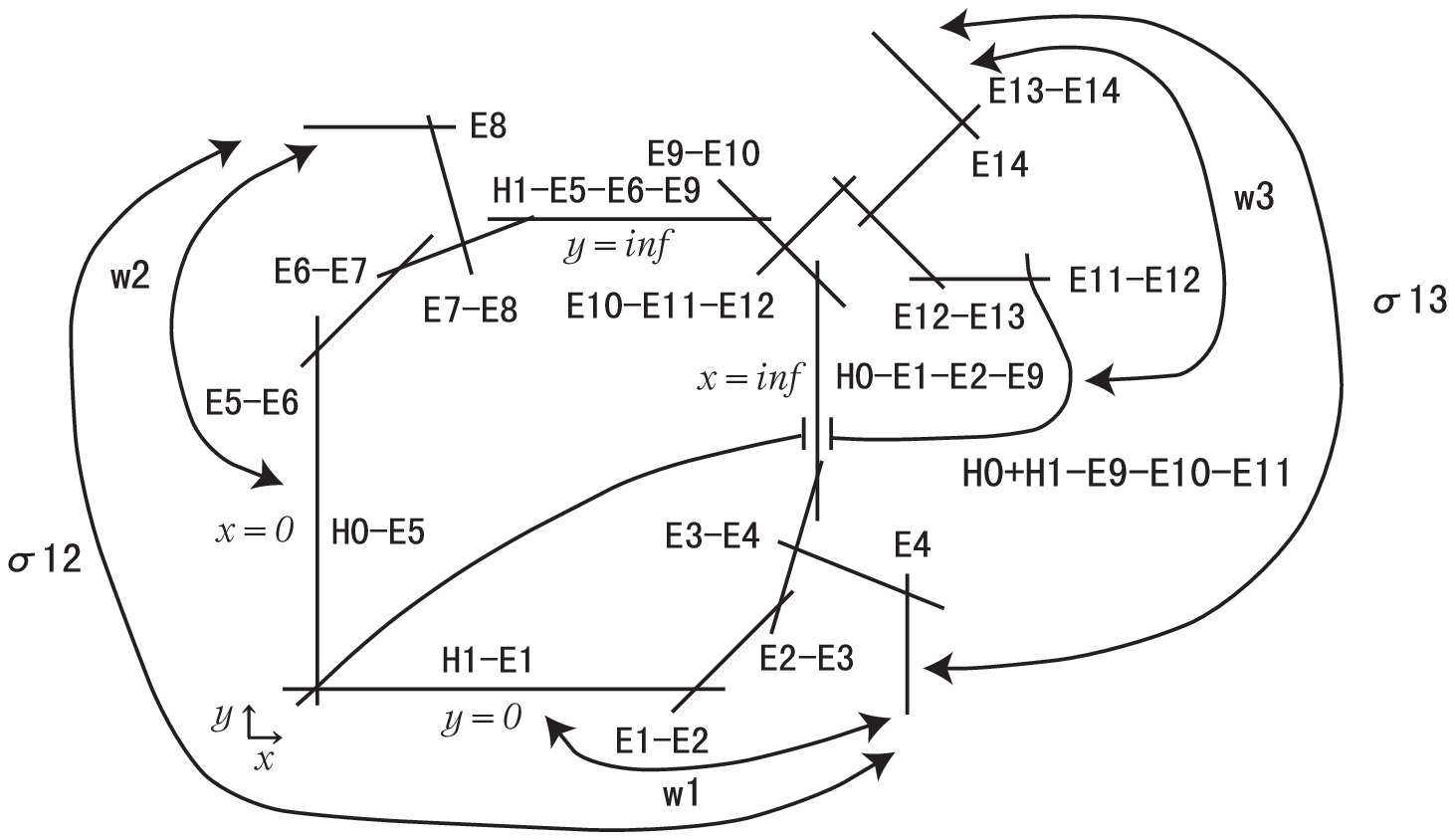}
\caption[]{}\label{divisors}
\end{center}
\end{figure}

The (linear equivalent classes of) prime divisors in 
(\ref{pexce}), (\ref{pline}) as elements of Pic($X$)  are described 
as\\ 
$C_0=E_4,~C_1=E_8,~C_2=E_{14},~C_3=H_0-E_5,~C_4=H_1-E_1 ~~(-1
\mbox{ curve)}$\\
$D_0= E_1-E_2,~ D_1= E_2-E_3,~ D_2=E_3-E_4,$\\
$D_3=E_5-E_6,~ D_4=E_6-E_7,~D_5=E_7-E_8,$\\
$D_6=E_9-E_{10},~D_7=E_{11}-E_{12},
~D_8=E_{12}-E_{13}, ~D_9=E_{13}-E_{14}~~(-2 \mbox{ curve)} $\\
$D_{10}=H_0-E_1-E_2-E_9,~ D_{11}=H_1-E_5-E_6-E_9,$\\
$D_{12}=E_{10}-E_{11}-E_{12}~~(-3 \mbox{ curve)}$\\
where by $n$ curve we mean a curve whose self-intersection number is $n$.
See Fig.\ref{divisors}.

The anti-canonical divisor $-K_X$ can be reduced uniquely (see appendix A) 
to prime divisors as
\begin{eqnarray}
&&D_0+2D_1+D_2+D_3+ 2D_4+D_5 \nonumber \\
&&+3D_6+D_7+2D_8+D_9+2D_{10}+2D_{11}+2D_{12}\label{dkx}
\end{eqnarray}
and the connection of $D_i$ are expressed by the following diagram.

\begin{eqnarray}\label{ddynkin}
\begin{picture}(400,100)
\put( 75,20){\line(1,0){20}}
\put(100,20){\line(1,0){20}}
\put(125,20){\line(1,0){20}}
\put(150,20){\line(1,0){20}}
\put(175,20){\line(1,0){20}}
\put(200,20){\line(1,0){20}}
\put(97.5,22.5){\line(0,1){20}}
\put(147.5,22.5){\line(0,1){20}}
\put(147.5,47.5){\line(0,1){20}}
\put(197.5,22.5){\line(0,1){20}}
\put(150,70){\line(1,0){20}}
\put(145,70){\line(-1,0){20}}
\put(72.5,20){\circle{5}}
\put(97.5,45){\circle{5}}
\put(97.5,20){\circle{5}}
\put(122.5,20){\circle*{5}}
\put(147.5,20){\circle{5}}
\put(172.5,20){\circle*{5}}
\put(197.5,20){\circle{5}}
\put(222.5,20){\circle{5}}

\put(147.5,45){\circle*{5}}
\put(147.5,70){\circle{5}}
\put(122.5,70){\circle{5}}
\put(172.5,70){\circle{5}}
\put(197.5,45){\circle{5}}
\put(72.5,10){\makebox(0,0){$D_3$}}
\put(97.5,10){\makebox(0,0){$D_4$}}
\put(122.5,10){\makebox(0,0){$D_{11}$}}
\put(147.5,10){\makebox(0,0){$D_6$}}
\put(172.5,10){\makebox(0,0){$D_{10}$}}
\put(197.5,10){\makebox(0,0){$D_{1}$}}
\put(222.5,10){\makebox(0,0){$D_0$}}
\put(112,45){\makebox(0,0){$D_5$}}
\put(162,45){\makebox(0,0){$D_{12}$}}
\put(212,45){\makebox(0,0){$D_2$}}
\put(122.5,80){\makebox(0,0){$D_9$}}
\put(147.5,80){\makebox(0,0){$D_8$}}
\put(172.5,80){\makebox(0,0){$D_7$}}

\put(280,50){\circle{5}}
\put(280,35){\circle*{5}}
\put(310,50){\makebox(0,0){$-2$ curve}}
\put(310,35){\makebox(0,0){$-3$ curve}}
\put(265,20){\line(1,0){20}}
\put(320,20){\makebox(0,0){intersection}}
\end{picture}\\
\end{eqnarray}

The HV eq.(\ref{hv}) acts on curves as (\ref{actd}).
Hence the HV eq. acts on Pic($X$) as
\begin{eqnarray*}%\label{actp}
\left(
\begin{array}{c}
H_0\\
H_1,~~E_1,~~ E_2\\
E_3,~~E_4,~~E_5,~~E_6\\
E_7,~~E_8,~~E_9,~~E_{10}\\
E_{11},~~E_{12},~~E_{13},~~E_{14}
\end{array}\right)
\to
\left( \begin{array}{c}
3H_0+H_1-E_5-E_6-E_7-E_8-E_9-E_{10}\\
H_0,~~H_0-E_8,~~H_0-E_7 \\
H_0-E_6,~~H_0-E_5,~~E_{11},~~E_{12}\\
E_{13},~~E_{14},~~H_0-E_{10},~~H_0-E_9\\
E_1,~~ E_2,~~E_3,~~E_4
\end{array}\right)
\end{eqnarray*}
(this table means $\overline{H_0}=3H_0+H_1-E_5-E_6-E_7-E_8-E_9-E_{10},$
$\overline{H_1}=H_0,$  $\overline{E_1}=H_0-E_8$ and so on)
and their linear combinations. \\ 

\remark
Let $\theta $ be an isomorphism from the rational surface $X$
to the rational surface $X'$.
Let $D$ be a divisor and $[D]$ its class.
The class of $\theta(D)$ coincides with the class 
$\theta([D])\in {\rm Pic}(X')$ and
the action of $\theta$ on Pic($X$) ($\cong$ Pic($X'$)) is linear. \\

\subsection{Cremona isometries and the root system}

\quad \\
{\bf Defnition.}
An automorphism $s$ of Pic($X$) is called a {\it Cremona isometry}
if the following three properties are satisfied: \\
a) $s$ preserves the intersection form in Pic($X$);\\
b) $s$ leaves $K_X$ fixed; \\
c) $s$ leaves the semigroup of effective classes of divisors invariant. \\

In general, if a birational mapping can be lifted to an isomorphism from
$X$ to $X'$ by blow-ups, its action on the resulting Picard group is 
always a Cremona isometry. We will show that the group of Cremona
isometries is an extended Weyl group of hyperbolic type. In the next section we
will show these Cremona isometries can be realized  as Cremona 
transformations, i.e. birational mappings, on 
${\mathbb P}^1\times {\mathbb P}^1$. 

\begin{lemma}\label{lemaut}
Let $s$ be a Cremona isometry, then\\
\noindent {\rm c')} $s$ is an automorphism 
of the diagram(\ref{ddynkin}). 
\end{lemma}

\bp
First we show that for any $i\in\{0,1,\cdots,12\}$ there exists 
$j\in\{0,\cdots,12\}$ such that $D_i=s(D_j)$.
Notice that $-K_X$ can be uniquely reduced to prime divisors 
in the form $-K_X=\sum_i  m_i D_i$ (see (\ref{dkx})) and 
the condition b). We have  
$s(-K_X)=-K_X=\sum_i  m_i s(D_i)$, where all $s(D_i)$ are effective
divisors due to the condition c) (and moreover 
$D_i\cdot D_i=s(D_i)\cdot s(D_i)$). By the uniqueness of decomposition 
of $-K_X$, we have that for any $i$ there exists $j$ such that
$D_i=s(D_j)$ and $m_i=m_j$. According to this fact and the condition a) we have the 
lemma. (Another proof is shown in \cite{looijenga,sakai}) \ep\\

Let us define $<D_i>$ and $<D_i>^{\bot}$ as  
$$<D_i>=\sum_{i=0}^{12}{\mathbb Z}D_i$$
and
$$<D_i>^{\bot}=\{\alpha \in {\rm Pic}(X); \alpha \cdot D_i=0  
\mbox{ for }i=0,1,\cdots,12\}.$$

\begin{lemma}\label{leminv}
The Cremona isometry $s$ leaves $<D_i>^{\bot}$ invariant.
\end{lemma}

\bp 
Let $\alpha\in <D_i>^{\bot}$. By the condition c') we have that 
for any $i\in\{0,\cdots,12\}$ there exists 
$j\in\{0,\cdots,12\}$ such that $s(\alpha)\cdot D_i= s(\alpha)\cdot s(D_j)
=\alpha \cdot D_j=0$.
It implies $s(\alpha)\in <D_i>^{\bot}$. \ep \\

In this case $<D_i>^{\bot}$ can be written as
$$<D_i>^{\bot}=<\alpha_i>:={\mathbb Z}\alpha_1+{\mathbb Z}\alpha_2+
{\mathbb Z}\alpha_3$$ 
where 
\begin{eqnarray*}
\alpha_1&=&2H_1-E_1-E_2-E_3-E_4,\nonumber\\
\alpha_2&=&2H_0-E_5-E_6-E_7-E_8,\\
\alpha_3&=&2H_0+2H_1-2E_9-2E_{10}-E_{11}-E_{12}-E_{13}-E_{14}.\nonumber
\end{eqnarray*}

We consider $<\alpha_i>$ with the intersection form 
to be a root lattice with a symmetric bilinear form. 
Let us define the transformation $w_i (i=1,2,3)$ on $<\alpha_i>$ as
\begin{eqnarray}\label{wrbot}
w_i(\alpha)=\alpha-2\frac{\alpha_i\cdot \alpha}{\alpha_i\cdot \alpha_i}\alpha_i
\end{eqnarray}
for $\alpha \in <\alpha_i>$.
The transformation 
$w_i(\alpha_j)$ has the form $w_i(\alpha_j)=\alpha_j- c_{ij}\alpha_i$,
where $c_{ij}= 2(\alpha_i\cdot \alpha_j)/(\alpha_i\cdot \alpha_i)$,
and the matrix $c_{ij}$ becomes a generalized Cartan matrix.
Here, the generalized Cartan matrix and its Dynkin diagram are 
of the hyperbolic type $H_{71}^{(3)}$ \cite{wan} as follows:
\begin{eqnarray}\left[
\begin{array}{rrr}
2&-2&-2\\-2&2&-2\\-2&-2&2
\end{array}
\right]
\quad &\mbox{ and }&
\begin{picture}(0,25)(-20,30)
\put(-2.5,20){\circle{5}}
\put(-2.5,10){\makebox(0,0){$\alpha_1$}}
\put(42.5,20){\circle{5}}
\put(42.5,10){\makebox(0,0){$\alpha_2$}}
\put(20,42.5){\circle{5}}
\put(20,50){\makebox(0,0){$\alpha_3$}}
\put(2,19){\line(1,0){36}}
\put(2,21){\line(1,0){36}}
\put(1,24){\line(1,1){16}}
\put(-1,25){\line(1,1){16}}
\put(39,24){\line(-1,1){16}}
\put(41,25){\line(-1,1){16}}
\put(0,20){\line(2,1){10}}
\put(0,20){\line(2,-1){10}}
\put(-1,23){\line(1,6){2}}
\put(-1,23){\line(2,1){10}}
\put(17,41){\line(-1,0){10}}
\put(17,41){\line(-1,-6){2}}
\put(23,41){\line(1,0){10}}
\put(23,41){\line(1,-6){2}}
\put(40,20){\line(-2,-1){10}}
\put(40,20){\line(-2,1){10}}
\put(41,23){\line(-1,6){2}}
\put(41,23){\line(-2,1){10}}
\end{picture}\\
\end{eqnarray}
\medskip

Hence the group $W$ generated by the actions $w_1,w_2,w_3$,
is a Weyl group of hyperbolic type. The extended 
(including the full automorphism group
of the Dynkin diagram) Weyl group, $\widetilde{W}$, is generated by 
\begin{eqnarray}\label{weyl}
\{w_1,w_2,w_3,\sigma_{12},\sigma_{13}\}
\end{eqnarray}
and the fundamental relations: 
\begin{eqnarray}
&w_i^2=\sigma_{1j}^2=1,~~(\sigma_{12}\sigma_{13})^3=1,&\nonumber\\
&\sigma_{12}w_1=w_2\sigma_{12},~~\sigma_{12}w_2=w_1\sigma_{12},
~~\sigma_{12}w_3=w_3\sigma_{12}& \label{frel} \\
&\sigma_{13}w_1=w_3\sigma_{13},~~\sigma_{13}w_2=w_2\sigma_{13},~~
\sigma_{13}w_3=w_1\sigma_{13}&  \nonumber
\end{eqnarray}
where the action of $\sigma_{12}$ or $\sigma_{13}$ on $<\alpha_i>$
is defined by the exchange of indices of $\alpha_i$; 
the action of $\sigma_{1j}$ and $w_k$ on $\alpha_i$ can be summarized as follows:
\begin{eqnarray*}%\label{acta}
&\sigma_j(\alpha_i) \mbox{ and } w_k(\alpha_i)&\nonumber\\
&\begin{array}{|c|c|c|c|c|c|}\hline
        &\sigma_{12}&\sigma_{13}&w_1&w_2&w_3\\ \hline
\alpha_1\mapsto&\alpha_2&\alpha_3&-\alpha_1&\alpha_1+2\alpha_2
             &\alpha_1+2\alpha_3\\\hline
\alpha_2\mapsto&\alpha_1&\alpha_2&\alpha_2+2\alpha_1&-\alpha_2
             &\alpha_2+2\alpha_3\\\hline
\alpha_3\mapsto&\alpha_3&\alpha_1&\alpha_3+2\alpha_1
             &\alpha_3+2\alpha_2&-\alpha_3\\\hline
\end{array}&.
\end{eqnarray*}

Moreover, by the following property we have the fact that
the group of Cremona isometries  
is included in $\pm \widetilde{W}$. 

\begin{proposition}{\bf (\cite{kac} \S 5.10)}\label{pmw}
If the generalized Cartan Matrix $c_{ij}$ is a symmetric matrix
of finite, affine, or hyperbolic type, then the group of all automorphisms of 
$<\alpha_i>$ preserving the bilinear form is $\pm \widetilde{W}$.
\end{proposition}

\remark
If $s$ is a Cremona isometry, then
$-s$ can not satisfy the condition c). \\ 

Next we consider uniqueness for the extension of action of $\widetilde{W}$
to the action on Pic($X$).

\begin{lemma}\label{lemuni}
Let $s$ and $s'$ be Cremona isometries such that
the action of $s$ is identical to the action of $s'$ on $<\alpha_i>$,
then $s=s'$ as Cremona isometries, i.e. $s$ is identical to $s'$ 
as an automorphism of $\pic(X)$.
\end{lemma}

\bp
Let $s$ and $s'$ be Cremona isometries such that
the actions of $s$ is identical to the action of $s'$ 
on $<\alpha_i>$. The actions of $s'\circ s^{-1}$ on $<\alpha_i>$
is the identity. 

We investigate where the exceptional divisor $E_4$ is moved by the action of 
$s'\circ s^{-1}$. In $\{D_i~;i=0,\cdots,12 \}$, 
only $D_2$ has an intersection with $E_4$. By the condition c'),  
$s'\circ s^{-1}(D_2)$ is $D_0,D_2, D_3,D_5,D_7$ or $D_9$ 
and only $s'\circ s^{-1}(D_2)$ has an intersection with $s'\circ s^{-1}(E_4)$
in $\{s'\circ s^{-1}(D_i);~i=0,\cdots,12\} $.

i) Assume $s'\circ s^{-1}(D_2)=D_2$. $s'\circ s^{-1}(E_4)$ has an intersection
only with $D_2$ in $\{D_i~;i=0,\cdots,12 \}$ (this condition on the 
coefficients of basis of Pic($X$) can be considered
to be a system of linear equations of order $13$). Then we have
the general solution with three integers
$z_1,z_2,z_3$:
$$s'\circ s^{-1}(E_4)=E_4+z_1\alpha_1+z_2\alpha_2+z_3\alpha_3.$$
Multiplying this equation by $s'\circ s^{-1}(\alpha_i)=\alpha_i$, we have
the system of linear equations:
\begin{eqnarray*}
\left\{ \begin{array}{ccc}
1-4 z_1+4z_2+4z_3&=&1\\
0+4 z_1-4z_2+4z_3&=&0\\
0+4 z_1+4z_2-4z_3&=&0
\end{array}\right. .
\end{eqnarray*}
It implies $s'\circ s^{-1}(E_4)=E_4$.

ii) Assume $s'\circ s^{-1}(D_2)=D_0$. We have  
$s'\circ s^{-1}(E_4)= (H_1-E_1) +z_1\alpha_1+z_2\alpha_2+z_3\alpha_3$. 
Multiplying this equation by $s'\circ s^{-1}(\alpha_i)=\alpha_i$,
one has that this equation does not have integer solutions.

iii) The other cases. $s'\circ s^{-1}(D_2)=D_3,D_5,D_7$ or $D_9$
implies $s'\circ s^{-1}(E_4)= L +z_1\alpha_1+z_2\alpha_2+z_3\alpha_3$, 
where $L=H_0-E_5,E_8,H_0+H_1-E_9-E_{10}-E_{11}$ or $E_{14}$ respectively.
This implies that this equation does not have integer solutions.

According to i),ii) and iii) $s'\circ s^{-1}(E_4)=E_4$ and 
$s'\circ s^{-1}(D_2)=D_2$.

Analogously we have  $s'\circ s^{-1}(H_1-E_1)=H_1-E_1$ and 
$s'\circ s^{-1}(D_0)=D_0$ and so on. Due to this fact and the 
condition c'), $s'\circ s^{-1}$ must be the identity as an Cremona
isometry. This implies the lemma. \ep\\  

Next we consider the extension of actions of elements of  
$\widetilde{W}$ on $<\alpha_i>$
to the actions on Pic($X$). 
Let us define $\alpha_{i,j} ~(i=1,2,3~j=1,2)$ as
\begin{eqnarray*}
\begin{array}{ll}
\alpha_{1,1}=H_1-E_1-E_4, & \alpha_{1,2}=H_1-E_2-E_3,\\ 
\alpha_{2,1}=H_0-E_5-E_7, & \alpha_{2,2}=H_0-E_6-E_7,\\ 
\alpha_{3,1}=H_0+H_1-E_9-E_{10}-E_{11}-E_{14},&\\ 
\alpha_{3,2}=H_0+H_1-E_9-E_{10}-E_{12}-E_{13}&
\end{array}
\end{eqnarray*}
and define the action of  
$\alpha_{i,j} ~(i=1,2,3~j=1,2)$ on $\alpha\in <\alpha_i>$ as
$$w_{i,j}(\alpha):=\alpha- 
2\frac{\alpha_{i,j}\cdot \alpha}{\alpha_{i,j}\cdot\alpha_{i,j}}\alpha_{i,j} \quad.$$

It is easy to see that $w_i(\alpha)=w_{i,1}\circ w_{i,2}(\alpha)=
w_{i,2}\circ w_{i,1}(\alpha)$.
We define the action of $w_i$ on $D\in $Pic($X$) as 
$w_i(D):=w_{i,1}\circ w_{i,2}(D)=w_{i,2}\circ w_{i,1}(D)$. 
These actions are explicitly written as follows 
(See Fig.\ref{divisors}). (For brevity we did not write 
the invariant elements under each action.) 

{\small
\begin{eqnarray}\label{weyld}
\begin{array}{lll}
&w_1:&\left(\begin{array}{c}
H_0,\\
E_1,~~E_2,~~E_3,~~E_4
\end{array}\right)
\to
\left(\begin{array}{c}
H_0+2H_1-E_1-E_2-E_3-E_4\\
H_1-E_4,~~H_1-E_3,~~H_1-E_2,~~H_1-E_1
\end{array}\right)\\
&w_2:&\left(\begin{array}{c}
H_1,\\
E_5,~~E_6,~~E_7,~~E_8
\end{array}\right)
\to
\left(\begin{array}{c}
2H_0+H_1-E_5-E_6-E_7-E_8\\
H_0-E_8,~~H_0-E_7,~~H_0-E_6,~~H_0-E_5
\end{array}\right)\\
&w_3:&\left(\begin{array}{c}
H_0,~H_1,~E_9,~E_{10}\\
E_{11},~E_{12},~E_{13},~E_{14}
\end{array}\right)
\to
\left(\begin{array}{c}
H_0+\alpha_3,~H_1+\alpha_3,~E_9+\alpha_3,~E_{10}+\alpha_3\\
E_{11}+\alpha_{3,1},~E_{12}+\alpha_{3,2},~E_{13}+\alpha_{3,2},
~E_{14}+\alpha_{3,1}
\end{array}\right)
\end{array}
\end{eqnarray}}
We define the action of $\sigma_{12}$ and $\sigma_{13}$ on Pic($X$)
as follows.
{\small
\begin{eqnarray}\label{weyls}
\begin{array}{lll}
&\sigma_{12}:&\left(\begin{array}{c}
H_0,~~H_1,~~E_1,~~E_2,~~E_3\\
E_4,~~E_5,~~E_6,~~E_7,~~E_8
\end{array}\right)
\to
\left( \begin{array}{c}
H_1,~~H_0,~~E_5,~~E_6,~~E_7\\
E_8,~~E_1,~~E_2,~~E_3,~~E_4
\end{array}\right)\\
&\sigma_{13}:&\left(\begin{array}{c}
H_1,~~E_1,~~E_2\\
E_3,~~E_4,~~E_9,~~E_{10}\\
E_{11},~~E_{12}~~E_{13},~~E_{14}
\end{array}\right)
\to
\left( \begin{array}{c}
H_0+H_1-E_9-E_{10},~~E_{11},~~E_{12}\\
E_{13},~~E_{14},~~H_0-E_{10},~~H_0-E_9\\
E_1,~~E_2,~~E_3,~~E_4
\end{array}\right)\\
\end{array}
\end{eqnarray}}

By direct calculation, it is easy to check that each $w_i$ (or $\sigma_{1i}$)
expressed by 
(\ref{weyld}) (or (\ref{weyls}) resp.) acts on $<\alpha_i>$ as (\ref{wrbot})
(or as the exchanges of indices of $\alpha_i$ resp.)
and that they satisfy the fundamental
relations (\ref{frel}) 
(the later property is of course guaranteed by the uniqueness
of extension of $\widetilde{W}$). Moreover it is also easy to check that 
the actions of all elements of $\widetilde{W}$ on Pic($X$) satisfy the 
conditions a),b) and c').
   
\begin{theorem}
The group of Cremona isometries of $X$ is isomorphic to $\widetilde{W}$,
where $\widetilde{W}$ is generated by $\{w_1,w_2,w_3,\sigma_{12},\sigma_{13}\}$
and the fundamental relations (\ref{frel}). The actions of elements 
of $\widetilde{W}$ on $\pic(X)$ are given by (\ref{weyld}) and (\ref{weyls}) and their composition. 
\end{theorem}

  To show this theorem, it is enough to show that (\ref{weyld}) and 
(\ref{weyls}) satisfy the condition c). 
For this purpose, it is enough to realize
them as isomorphisms from  $X$ to $X'$, where $X$ and $X'$ have the same
semigroup of classes of effective divisors. 
We show this fact in the next subsection.

From  (\ref{weyld}) and (\ref{weyls}) it is straightforward to show that
the action of the HV eq. on Pic($X$) is identical
to the action of $w_2\circ \sigma_{13}\circ\sigma_{12}$.

\begin{corollary}
There does not exist any Cremona isometry of $X$ whose action on $\pic(X)$ 
commutes with the action of the HV eq.  
except  $(w_2\circ \sigma_{13}\circ \sigma_{12})^m$, where $m\in{\mathbb Z}$.
\end{corollary}

\bp
In this proof we denote $\sigma_{12}$ or $\sigma_{13}$ by 
$\sigma_2$ or $\sigma_3$ respectively and omit the symbol of composition 
$\circ$.
Each element of $\widetilde{W}$ can be uniquely
written in the form 
$$w_{i_1}w_{i_2}\cdots w_{i_n}s$$
where all indices of 
$w$ (or $\sigma$) are considered in Mod 3 (or 2 resp.) and $i_l\neq i_{l+1}$
and $s\in\{1,\sigma_j,\sigma_j\sigma_{j+1},\sigma_j\sigma_{j+1}\sigma_j\}$.
Assume $g=w_{i_1}w_{i_2}\cdots w_{i_n}s$ commutes with
$w_2\sigma_3\sigma_2$.

i) The case of $s=1$. According to the relation
$$w_2\sigma_3\sigma_2w_{i_1}w_{i_2}\cdots w_{i_n}=
w_{i_1}w_{i_2}\cdots w_{i_n}w_2\sigma_3\sigma_2,$$
we have the relation
$$w_2w_{i_1+1}w_{i_2+1}\cdots w_{i_n+1}\sigma_3\sigma_2=
w_{i_1}w_{i_2}\cdots w_{i_n}w_2\sigma_3\sigma_2.$$
It implies $i_1\equiv 2,i_2\equiv 3,\cdots,i_{n}\equiv n+1,2\equiv n+2$
and therefore there exists the integer $m$ such that $n=3m$.  
On the other hand
$(w_2\sigma_3\sigma_2)^{3m}=w_2w_3\cdots w_{n+1}$.
It implies $g=(w_2\sigma_3\sigma_2)^{3m}$.

ii) The case of 
$s=\sigma_3\sigma_2$ or $\sigma_2\sigma_3$. Similar to the case i),
$n$ must be $n=3m+1$ or $n=3m+2$ respectively and
$g$ becomes $(w_2\sigma_3\sigma_2)^{n}$.

iii) The case of $s=\sigma_j$. Suppose $j=2$.
According to the relation
$$w_2\sigma_3\sigma_2 w_{i_1}w_{i_2}\cdots w_{i_n}\sigma_2=
w_{i_1}w_{i_2}\cdots w_{i_n}\sigma_2w_2\sigma_3\sigma_2,$$
we have the relation
$$w_2w_{i_1+1}w_{i_2+1}\cdots w_{i_n+1}\sigma_3=
w_{i_1}w_{i_2}\cdots w_{i_n}w_1\sigma_2\sigma_3\sigma_2.$$
It implies $\sigma_3=\sigma_2\sigma_3\sigma_2$ but this is a contradiction.
Similarly $s=\sigma_3$ is impossible.

v) The case $s=\sigma_j\sigma_{j+1}\sigma_j$. 
Suppose $j=2$. Similar to the case iii), we have
the relation
$$w_2w_{i_1+1}w_{i_2+1}\cdots w_{i_n+1}\sigma_2=
w_{i_1}w_{i_2}\cdots w_{i_n}w_3\sigma_2\sigma_3\sigma_2\sigma_3\sigma_2.$$
It implies $\sigma_2=\sigma_2\sigma_3\sigma_2\sigma_3\sigma_2$ and hence
$1=\sigma_3\sigma_2\sigma_3\sigma_2$ which leads to a contradiction.
Similarly it can be shown that $s=\sigma_3\sigma_2\sigma_3$ is impossible.\ep\\

\noindent{\it Conclusion of the section.}\\
We have shown in this section that 
a) the ``Dynkin diagram''(\ref{ddynkin}) of the irreducible components of the 
anti-canonical divisor $-K_X$  does not correspond to a generalized
 Cartan matrix,
b) the group of Cremona isometries is isomorphic to an extended Weyl group
of hyperbolic type and 
c) there does not exist a Cremona isometry which is commutative  with 
the action of the HV eq. except itself, 
while for the discrete Pailev\'e equations a) affine type, b) affine type
and c) there does exist such a Cremona isometries except for the 
trivial exceptions
(the type of root system is $A_1^{(1)}$ etc.).

\section{The inverse problem}

A birational mapping is called a Cremona transformation.
One method for obtaining a Cremona transformation such that 
its action on Pic($X$) is a Cremona isometry 
is to interchange the blow down structures,
i.e. to interchange the procedure of blow downs.
Following this method, we construct the Cremona transformations
on ${\mathbb P}^1\times{\mathbb P}^1$
which yield the extended Weyl group (\ref{weyl}), (\ref{frel}) and thereby 
recover the HV eq.
from its action on Pic($X$).  

An element of $\widetilde{W}$  is an automorphism 
of Pic($X$) but does not have to be an automorphism of
$X$ itself, i.e. the blow-up points can be changed with a transformation 
satisfying the condition a),b) and c) in Section~3.2.
In order to do this, one has to consider not only autonomous
but also non-autonomous mappings. 

By $a_0,a_1,a_2,a_3,a_4,a_5,a_6$ or $a_7$  we denote the point of 
the $10,3,4,7,8,11,13,14$-th blow-up or the corresponding coordinates
and we call them ``the parameters''. 
In short these points can be
expressed as follows:
\begin{eqnarray*}
&&\left(\frac{1}{xy},xy^2\right)=(0,a_1), \quad
\left(\frac{1}{xy}, xy(xy^2-a_1)\right)=(0,a_2),\noindent\\
&&\left(x^2 y,\frac{1}{xy}\right)=(a_3,0)), \quad
\left(xy(x^2 y-a_3),\frac{1}{xy}\right)=(a_4,0),\noindent\\
&&\left(\frac{1}{x},\frac{x}{y}\right)=(0,a_0) ), \quad
\mbox{where we normalize }a_0 \mbox{ to be } a_0=1, \label{para}\\
&&\left(\frac{1}{x},x(\frac{x}{y}-1)\right)=(0,a_5), \quad
\left(\frac{1}{x(x(\frac{x}{y}-1)-a_5)},  
x(x(\frac{x}{y}-1)-a_5)^2 \right)=(0,a_6),\noindent\\
&&\left(\frac{1}{x(x(\frac{x}{y}-1)-a_5)}, 
x(x(\frac{x}{y}-1)-a_5)\left\{x(x(\frac{x}{y}-1)-a_5)^2-a_6 \right\} \right)=(0,a_7),\noindent
\end{eqnarray*}
where
$a_i\in {\mathbb C}$ and $a_1,a_3,a_6$ are nonzero.

The point of the $2,6,12$-th blow-up is determined by intersection numbers.
Moreover the point of the $1,5,9,10,11$-th blow-up can 
be fixed by acting with a suitable
automorphism of ${\mathbb P}^1 \times {\mathbb P}^1$, i.e. 
a M\"{o}bius transformation of each coordinate combination with an
exchange of 
the coordinates. We call this operation ``normalization''.
It can also be seen that we can normalize $a_5=1$ except the case $a_5=0$. 

In this section we consider a realization of the generating elements of 
$\widetilde{W}$ as Cremona transformations which can be lifted to 
isomorphisms from $X$ to $\ol{X}$,
where $\ol{X}$ is the same rational surface as $X$ except for a difference in
parameters. 

First we realize 
the action of $w_2$ as a Cremona transformation on 
${\mathbb P}^1\times {\mathbb P}^1$

\subsection{The calculation of interchanging the blow down structure}

In the following we shall present a scheme in which the blow down
structure is changed. This method is based on the following fact:

By $F_n$ we denote the $n$th Hirzebruch surface with the coordinate system
\begin{eqnarray}\label{hirz}
(\bigcirc, \triangle)\cup (\bigcirc, \frac{1}{\triangle})
\cup(\frac{1}{\bigcirc}, \bigcirc^n \triangle)
\cup( \frac{1}{\bigcirc}, \frac{1}{\bigcirc^n \triangle}).
\end{eqnarray}
Blowing up the $n$th Hirzebruch surface at the point
$(1/\bigcirc, \bigcirc^n \triangle)=(0,0)$
and blowing down along the line $1/\bigcirc=1/(\bigcirc^{n+1} \triangle)=0$,
we obtain the $n+1$-th Hirzebruch surface as follows:
\begin{eqnarray*}
\begin{array}{ccccc}
\vspace{3pt}&(\bigcirc, \triangle)&\cup (\bigcirc, \frac{1}{\triangle})
&\cup(\frac{1}{\bigcirc}, \bigcirc^n \triangle)
&\cup( \frac{1}{\bigcirc}, \frac{1}{\bigcirc^n \triangle})\\
\vspace{3pt}
\mapleft{{\rm up}}&(\bigcirc, \triangle)&\cup (\bigcirc, \frac{1}{\triangle})
&\cup(\frac{1}{\bigcirc}, \bigcirc^{n+1} \triangle)
\cup (\frac{1}{\bigcirc^{n+1} \triangle}, \bigcirc^n \triangle) 
& \cup( \frac{1}{\bigcirc}, \frac{1}{\bigcirc^n \triangle})\\
\mapright{{\rm down}}&
(\bigcirc, \triangle)&\cup (\bigcirc, \frac{1}{\triangle})
&\cup(\frac{1}{\bigcirc}, \bigcirc^{n+1} \triangle)
&\cup( \frac{1}{\bigcirc}, \frac{1}{\bigcirc^{n+1} \triangle}).
\end{array}
\end{eqnarray*}
On the other hand,
blowing up the $n$th Hirzebruch surface at the point\\
$(1/\bigcirc, 1/(\bigcirc^n \triangle))=(0,0)$
and blowing down along the line $1/\bigcirc=\bigcirc^{n-1} \triangle=0$,
we obtain the $n-1$-th Hirzebruch surface as follows:
\begin{eqnarray*}
\begin{array}{ccccc}
\vspace{3pt}
&(\bigcirc, \triangle)&\cup (\bigcirc, \frac{1}{\triangle})
&\cup(\frac{1}{\bigcirc}, \bigcirc^n \triangle)
&\cup( \frac{1}{\bigcirc}, \frac{1}{\bigcirc^n \triangle})\\
\vspace{3pt}
\mapleft{{\rm up}}&(\bigcirc, \triangle)&\cup (\bigcirc, \frac{1}{\triangle})
&\cup(\frac{1}{\bigcirc}, \bigcirc^n \triangle)
&\cup( \frac{1}{\bigcirc}, \frac{1}{\bigcirc^{n-1} \triangle})
\cup(\bigcirc^{n-1} \triangle, \frac{1}{\bigcirc^n \triangle})\\
\mapright{{\rm down}}&(\bigcirc, \triangle)&
\cup (\bigcirc, \frac{1}{\triangle})
&\cup(\frac{1}{\bigcirc}, \bigcirc^{n-1} \triangle)
&\cup( \frac{1}{\bigcirc}, \frac{1}{\bigcirc^{n-1} \triangle}).
\end{array}
\end{eqnarray*}

The Next figure shows the order of the blow-ups and the blow downs
to obtain the Cremona transformation corresponding to $w_2$.
(This table has to be read from the left to the right.)

\begin{picture}(400,180)(100,0)
\put(145,112.5){\line(0,1){50}}
\put(100,122.5){\line(1,0){50}}
\put(105,112.5){\line(0,1){50}}
\put(100,152.5){\line(1,0){50}}
\put(108,105){\makebox(0,0){$H_0$}}
\put(108,95){\makebox(0,0){$x=0$}}
\put(152,105){\makebox(0,0){$H_0$}}
\put(152,95){\makebox(0,0){$x=\infty$}}
\put(170,152.5){\makebox(0,0){$H_1$}}
\put(170,142.5){\makebox(0,0){$y=\infty$}}

\put(135,65){\vector(-3,4){14}}
\put(175,12.5){\line(0,1){50}}
\put(130,22.5){\line(1,0){50}}
\put(135,12.5){\line(0,1){30}}
\put(136,12.5){\line(0,1){30}}
\put(145,52.5){\line(1,0){35}}
\put(130,32){\line(1,1){30}}
\put(138,5){\makebox(0,0){$H_0-E_5$}}
\put(182,5){\makebox(0,0){$H_0$}}
\put(210,52.5){\makebox(0,0){$H_1-E_5$}}
\put(160,70){\makebox(0,0){$E_5$}}

\put(185,65){\vector(3,4){14}}
\put(245,112.5){\line(0,1){50}}
\put(200,122.5){\line(1,0){50}}
\put(205,112.5){\line(0,1){50}}
\put(200,152.5){\line(1,0){50}}
\put(208,105){\makebox(0,0){$H_0$}}
\put(252,105){\makebox(0,0){$H_0$}}
\put(280,152.5){\makebox(0,0){$H_1-E_5$}}

\put(255,65){\vector(-3,4){14}}
\put(285,12.5){\line(0,1){50}}
\put(240,22.5){\line(1,0){50}}
\put(245,12.5){\line(0,1){30}}
\put(246,12.5){\line(0,1){30}}
\put(255,52.5){\line(1,0){35}}
\put(240,32){\line(1,1){30}}
\put(248,5){\makebox(0,0){$H_0-E_6$}}
\put(292,5){\makebox(0,0){$H_0$}}
\put(325,52.5){\makebox(0,0){$H_1-E_5-E_6$}}
\put(270,70){\makebox(0,0){$E_6$}}

\put(305,65){\vector(3,4){14}}
\put(365,112.5){\line(0,1){50}}
\put(320,122.5){\line(1,0){50}}
\put(325,112.5){\line(0,1){50}}
\put(320,152.5){\line(1,0){50}}
\put(328,105){\makebox(0,0){$H_0$}}
\put(372,105){\makebox(0,0){$H_0$}}
\put(405,152.5){\makebox(0,0){$H_1-E_5-E_6$}}

\put(395,65){\vector(-3,4){14}}
\put(425,12.5){\line(0,1){50}}
\put(380,22.5){\line(1,0){50}}
\put(385,12.5){\line(0,1){50}}
\put(386,12.5){\line(0,1){50}}
\put(380,52.5){\line(1,0){50}}
\put(360,37){\line(1,0){30}}
\put(388,5){\makebox(0,0){$H_0-E_7$}}
\put(432,5){\makebox(0,0){$H_0$}}
\put(465,52.5){\makebox(0,0){$H_1-E_5-E_6$}}
\put(360,27){\makebox(0,0){$E_7$}}
\put(435,65){\vector(3,4){14}}
\put(437,75){\makebox(0,0){*}}
\end{picture}\\

\begin{picture}(400,180)(50,0)

\put(110,75){\vector(3,4){14}}
\put(112,85){\makebox(0,0){*}}
\put(165,110){\line(0,1){50}}
\put(130,130){\line(3,2){40}}
\put(110,135){\line(1,0){40}}
\put(130,140){\line(3,-2){40}}
\put(110,125){\makebox(0,0){$H_0$}}
\put(167,100){\makebox(0,0){$H_0$}}
\put(205,160){\makebox(0,0){$H_0+H_1-E_5$}}
\put(216,148){\makebox(0,0){$-E_6-E_7$}}

\put(185,70){\vector(-3,4){14}}
\put(245,10){\line(0,1){50}}
\put(210,30){\line(3,2){40}}
\put(190,35){\line(1,0){40}}
\put(190,34){\line(1,0){40}}
\put(210,40){\line(3,-2){40}}
\put(170,60){\line(1,-1){30}}
\put(160,55){\makebox(0,0){$E_8$}}
\put(165,35){\makebox(0,0){$H_0-E_8$}}
\put(247,0){\makebox(0,0){$H_0$}}
\put(290,60){\makebox(0,0){$H_0+H_1-E_5$}}
\put(301,48){\makebox(0,0){$-E_6-E_7$}}

\put(260,75){\vector(3,4){14}}
\put(315,110){\line(0,1){50}}
\put(280,130){\line(3,2){40}}
\put(260,135){\line(1,0){40}}
\put(280,140){\line(3,-2){40}}
\put(260,125){\makebox(0,0){$H_0$}}
\put(317,100){\makebox(0,0){$H_0$}}
\put(360,160){\makebox(0,0){$2H_0+H_1-E_5$}}
\put(365,148){\makebox(0,0){$-E_6-E_7-E_8$}}
\end{picture}\\
where double lines mean the lines which are blown down in the next steps.   
\bigskip

The calculation of changing the blow down structure is as follows.
{\small
\begin{eqnarray*}
&&(x,y)\cup(x,\frac{1}{y})\cup(\frac{1}{x},y)\cup
(\frac{1}{x},\frac{1}{y}) ~~={\mathbb P}^1\times{\mathbb P}^1\\
&\leftarrow&(x,y)\cup(x,\frac{1}{xy})\cup(xy,\frac{1}{y})\cup(\frac{1}{x},y)
\cup(\frac{1}{x},\frac{1}{y})\\
&\rightarrow&(x,xy)\cup(x,\frac{1}{xy})\cup(\frac{1}{x},y)\cup
(\frac{1}{x},\frac{1}{y}) ~~=F_1\\
&\leftarrow&(x,xy)\cup(x,\frac{1}{x^2y})\cup(x^2y,\frac{1}{xy})\cup
(\frac{1}{x},y)\cup(\frac{1}{x},\frac{1}{y})\\
&\rightarrow&(x,x^2y)\cup(x,\frac{1}{x^2y})\cup(\frac{1}{x},y)\cup
(\frac{1}{x},\frac{1}{y}) ~~=F_2\\
&\sim&\quad(x,x^2y-a_3)\cup(x,\frac{1}{x^2y-a_3})\cup
(\frac{1}{x},\frac{x^2y-a_3}{x^2})\cup
(\frac{1}{x},\frac{x^2}{x^2y-a_3}) ~~=F_2\\
&\leftarrow&(x,\frac{x^2y-a_3}{x})\cup(\frac{x}{x^2y-a_3},x^2y-a_3)
\cup(x,\frac{1}{x^2y-a_3})\cup(\frac{1}{x},\frac{x^2y-a_3}{x^2})\cup
(\frac{1}{x},\frac{x^2}{x^2y-a_3}) \\
&\rightarrow&(x,\frac{x^2y-a_3}{x})
\cup(x,\frac{x}{x^2y-a_3})\cup(\frac{1}{x},\frac{x^2y-a_3}{x^2})\cup
(\frac{1}{x},\frac{x^2}{x^2y-a_3}) ~~=F_1\\
&\sim&\quad
(x,\frac{a_3(x^2y-a_3)-a_4x}{a_3x})
\cup(x,\frac{a_3x}{a_3(x^2y-a_3)-a_4x}) \cup \cdots
 ~~=F_1\\
&\leftarrow&(x,\frac{a_3(x^2y-a_3)-a_4x}{a_3x^2})\cup 
(\frac{a_3x^2}{(a_3(x^2y-a_3)-a_4x},\frac{a_3(x^2y-a_3)-a_4x}{a_3x})\\
&&\hspace{3cm} \cup(x,\frac{a_3x}{a_3(x^2y-a_3)-a_4x})\cup \cdots\\
&\rightarrow&(x,\frac{a_3(x^2y-a_3)_-a_4x}{a_3x^2})
\cup(x,\frac{a_3x^2}{a_3(x^2y-a_3)-a_4x})\cup \cdots
 ~~={\mathbb P}^1\times{\mathbb P}^1
\end{eqnarray*}
}
where $\cdots$ is
$$(\frac{1}{x},\frac{a_3(x^2y-a_3)-a_4x}{a_3x^2})\cup
(\frac{1}{x},\frac{a_3x^2}{a_3(x^2y-a_3)-a_4x})$$
and $\sim$ means an automorphism of the Hirzebruch surface
and is determined as the point of blow-up in (\ref{para}) is moved 
to the origin.

Writing 
$$w_2':(x,y)\mapsto (x,y-\frac{a_3}{x^2}-\frac{a_4}{a_3 x})$$
we obtain $w_2=t \circ w_2'$
where $t$ is an automorphism of ${\mathbb P}^1\times{\mathbb P}^1$.
By taking a suitable $t$, we can normalize $w_2$ to get
the required result.

\subsection{Normalization and the action on the space of parameters}

First we determine the automorphism for normalization $t$. 
By (\ref{weyld}), $w_2$ does not move the points
$(x,y)=(\infty,0),(\infty,\infty)$. 
According to the fact: 
$w_2': (\infty,0)\mapsto (\infty,0), (\infty,\infty)\mapsto (\infty,\infty)$,
$t$ is reduced to the mapping $t:(x,y)\mapsto (c_1 x+c_2, c_3 y)$,
where $c_1,c_2,c_3\in {\mathbb C}$ are nonzero constants. 

Similarly,  $w_2$ moves the point $a_3$ to the proper 
transform of the point of the $6$-th blow-up.
We denote this fact as  
$$\left.(\ol{u_6},\ol{v_6})\right|_{(u_6,v_6)=(a_3,0)}=(0,0) ,$$ 
where $(u_n,v_n)$ is the coordinate of the $n$th blow-up.
On the other hand,
\begin{eqnarray*}
&&\left.(\ol{u_6},\ol{v_6})\right|_{(u_6,v_6)=(a_3,0)}\\
&=&\left.\left(\bar{x}^2 \bar{y},\frac{1}{\bar{x}\bar{y}}\right)\right|_{(u_6,v_6)=(a_3,0)}\\
&=& \left.\left(x^2 (y-\frac{a_3}{x^2}-\frac{a_4}{a_3 x}),
\frac{a_3x}{a_3(x^2y-a_3)-a_4x}\right) \right|_{(u_6,v_6)=(a_3,0)}\\
&=& \left.\left(-a_3+u_6-\frac{a_4 u_6 v_6}{a_3},
- \frac{a_3 u_6 v_6}{a_3^2-a_3 u_6+a_4 u_6 v_6}\right)\right|_{(u_6,v_6)=(a_3,0)}
\\
&=&(0,0)
\end{eqnarray*}
holds. Hence $t$ does not move the point $(x,y)=(0,\infty)$ and therefore
$c_2=0$. 

Similarly, since $w_2$ does not move the points of 
the $10$-th and the $11$-th blow-ups, we have $c_1=c_3=1$
(moreover we can normalize $a_5$ to be $a_5=1$ or $a_5=0$
by taking a suitable value of $c_1=c_3$). 
Hence $t$ has to be 
the identity. 

Next we calculate how the parameters $a_1,a_2,a_3,a_4,a_6,a_7$ are 
changed by the action of $w_2$. Notice that $w_2$ is an isomorphism
from $X$ to $X'$, where $w_2$ satisfy the conditions a),b) and c) 
in Section~3.2. and therefore $X$ and $X'$ have the same sequence of 
blow-ups except their parameters. 
By $\overline{a_i}$ we denote the parameter of the $i$th blow-up of $X'$.
  
Since the action of $w_2$ moves the points of blow-ups as follows
\begin{eqnarray*}
(u_2,v_2)=(0,a_1) &\mapsto& 
(\overline{u_2},\overline{v_2})=(0,\overline{a_1})\\ 
(u_3,v_3)=(0,a_2) &\mapsto& 
(\overline{u_3},\overline{v_3})=(0,\overline{a_2})\\
(u_6,v_6)=(0,0) &\mapsto& 
(\overline{u_6},\overline{v_6})=(\overline{a_3},0)\\
(xy,1/y)=(0,0) &\mapsto& 
(\overline{u_7},\overline{v_7})=(\overline{a_4},0)\\
(u_{12},v_{12})=(0,a_6) &\mapsto& 
(\overline{u_{12}},\overline{v_{12}})=(0,\overline{a_6})\\ 
(u_{13},v_{13})=(0,a_7) &\mapsto& 
(\overline{u_{13}},\overline{v_{13}})=(0,\overline{a_7}), 
\end{eqnarray*}
$\overline{a_i}$ can be calculated. For example $\overline{a_1}$ 
is calculated as follows
\begin{eqnarray*}
(0,\overline{a_1})&=&\left.(\overline{u_2},\overline{v_2})
\right|_{(u_2,v_2)=(0,a_1)}\\
&=&\left.( \frac{1}{\bar{x}\bar{y}},\bar{x}\bar{y}^2)
\right|_{(u_2,v_2)=(0,a_1)}\\
&=&\left. \left(\frac{a_3u_2}{a_3 - a_4u_2 - a_3^2u_2^3v_2},
\frac{v_2(-a_3 + a_4u_2 + a_3^2u_2^3v_2)^2}{a_3^2}
\right)\right|_{(u_2,v_2)=(0,a_1)}\\
&=&(0,a_1),
\end{eqnarray*}
and therefore $\ol{a_1}=a_1$. 

Similarly we can calculate $\ol{a_2},\ol{a_3},\ol{a_4}$ as follows:
\begin{eqnarray*}
(0,\overline{a_2})&=& \left.( \frac{1}{\bar{x}\bar{y}},
\bar{x}\bar{y}(\bar{x}\bar{y}^2-\ol{a_1})
\right|_{(u_3,v_3)=(0,a_2)}\\
&=&(0,a_2-\frac{a_1a_4}{a_3}),
\end{eqnarray*}
\begin{eqnarray*}
(\overline{a_3},0)&=& \left.(\bar{x}^2\bar{y}, \frac{1}{\bar{x}\bar{y}})
\right|_{(u_6,v_6)=(0,0)}\\
&=&(-a_3,0),
\end{eqnarray*}
\begin{eqnarray*}
(\overline{a_4},0)&=& \left.( 
\bar{x}\bar{y}(\bar{x}^2\bar{y}-\ol{a_3}),\frac{1}{\bar{x}\bar{y}})
\right|_{(xy,1/y)=(0,0)}\\
&=&(a_4,0).
\end{eqnarray*}

Consequently $w_2$ changes the parameters $a_i$ as 
\begin{eqnarray*}
\begin{array}{llll}
\ol{a_1}=a_1,&\ol{a_2}=a_2- 2a_1a_4/a_3,&\ol{a_3}=-a_3,&\ol{a_4}=a_4,\\
\ol{a_5}=a_5, &\ol{a_6}=a_6, &\ol{a_7}=a_7+2a_4a_6/a_3&.
\end{array}
\end{eqnarray*}

We write the action of $w_2$ on ${\mathbb P}^1\times {\mathbb P}^1$ and 
the space of parameters together as
\begin{eqnarray*}
w_2:&&(x,y;a_1,a_2,a_3,a_4,a_5,a_6,a_7)\nonumber\\
&\mapsto& (\ol{x},\ol{y}~;~\ol{a_1},\ol{a_2},
\ol{a_3},\ol{a_4},\ol{a_5},\ol{a_6},\ol{a_7})\nonumber\\
&=&\left(x,y-\frac{a_3}{x^2}-\frac{a_4}{a_3x}~;~a_1,a_2- \frac{2a_1a_4}{a_3},
-a_3,a_4,a_5,a_6,a_7+\frac{2a_4a_6}{a_3}\right).
\end{eqnarray*}
Here, in the calculation of the next iteration step we have to use 
$\overline{a_3}=-a_3$ instead of $a_3$ and so on.

As was remarked before the mapping $w_2$ is of order 2 
as an element of an extended Weyl group and can be lifted to an isomorphism 
from $X$ to $\ol{X}$.

\subsection{The actions of other elements}

Next we calculate the action of $\sigma_{13}$ from $X$ to $\ol{X}$.

The following figure shows the order of the blow-ups and the blow downs.

\begin{picture}(400,180)(100,0)
\put(145,112.5){\line(0,1){50}}
\put(100,122.5){\line(1,0){50}}
\put(105,112.5){\line(0,1){50}}
\put(100,152.5){\line(1,0){50}}
\put(108,105){\makebox(0,0){$H_0$}}
\put(108,95){\makebox(0,0){$x=0$}}
\put(152,105){\makebox(0,0){$H_0$}}
\put(152,95){\makebox(0,0){$x=\infty$}}
\put(170,152.5){\makebox(0,0){$H_1$}}
\put(170,142.5){\makebox(0,0){$y=\infty$}}

\put(135,65){\vector(-3,4){14}}
\put(175,12.5){\line(0,1){30}}
\put(174,12.5){\line(0,1){30}}
\put(130,22.5){\line(1,0){50}}
\put(135,12.5){\line(0,1){50}}
\put(130,52.5){\line(1,0){35}}
\put(180,32){\line(-1,1){30}}
\put(138,5){\makebox(0,0){$H_0$}}
\put(182,5){\makebox(0,0){$H_0-E_9$}}
\put(190,52.5){\makebox(0,0){$H_1-E_9$}}
\put(160,70){\makebox(0,0){$E_9$}}

\put(185,65){\vector(3,4){14}}
\put(245,112.5){\line(0,1){50}}
\put(200,122.5){\line(1,0){50}}
\put(205,112.5){\line(0,1){50}}
\put(200,152.5){\line(1,0){50}}
\put(208,105){\makebox(0,0){$H_0$}}
\put(252,105){\makebox(0,0){$H_0$}}
\put(280,152.5){\makebox(0,0){$H_1-E_9$}}

\put(255,65){\vector(-3,4){14}}
\put(285,12.5){\line(0,1){50}}
\put(284,12.5){\line(0,1){50}}
\put(240,22.5){\line(1,0){50}}
\put(245,12.5){\line(0,1){50}}
\put(240,52.5){\line(1,0){50}}
\put(280,35.5){\line(1,0){30}}
\put(248,5){\makebox(0,0){$H_0$}}
\put(292,5){\makebox(0,0){$H_0-E_{10}$}}
\put(315,52.5){\makebox(0,0){$H_1-E_9$}}
\put(320,27){\makebox(0,0){$E_{10}$}}

\put(305,65){\vector(3,4){14}}
\put(325,112.5){\line(0,1){50}}
\put(360,132.5){\line(-3,2){40}}
\put(360,142.5){\line(-3,-2){40}}
\put(340,137.5){\line(1,0){40}}
\put(328,105){\makebox(0,0){$H_0$}}
\put(380,127.5){\makebox(0,0){$H_0$}}
\put(385,152.5){\makebox(0,0){$H_0+H_1-E_9-E_{10}$}}
\end{picture}\\

\bigskip

Its calculation is as follows.
{\small
\begin{eqnarray*}
&&(x,y)\cup(x,\frac{1}{y})\cup(\frac{1}{x},y)\cup
(\frac{1}{x},\frac{1}{y}) ={\mathbb P}^1\times{\mathbb P}^1\\
&\leftarrow&(x,y)\cup(x,\frac{1}{y})\cup(\frac{1}{x},y)
\cup(\frac{1}{x},\frac{x}{y}) \cup(\frac{y}{x},\frac{1}{y})\\
&\rightarrow&(x,y)\cup(x,\frac{1}{y})\cup(\frac{1}{x},\frac{y}{x})\cup
(\frac{1}{x},\frac{x}{y}) =F_1\\
&\sim&
(x,y-x)\cup(x,\frac{1}{y-x})\cup(\frac{1}{x},\frac{y-x}{x})\cup
(\frac{1}{x},\frac{x}{y-x}) =F_1\\
&\leftarrow&(x,y-x)\cup(x,\frac{1}{y-x})\cup(\frac{1}{x},y-x)
\cup(\frac{1}{y-x},\frac{y-x}{x})
\cup(\frac{1}{x},\frac{x}{y-x})\\
&\rightarrow&(x,y-x)\cup(x,\frac{1}{y-x})\cup(\frac{1}{x},y-x)
\cup(\frac{1}{x},\frac{1}{y-x})={\mathbb P}^1\times{\mathbb P}^1
\end{eqnarray*}
}

Similar to the case of $w_2$, we have the action of $\sigma_{13}$
on $X$ and the space of parameters as follows
\begin{eqnarray*}
\sigma_{13}:&&(x,y;a_1,a_2,a_3,a_4,a_5,a_6,a_7)\nonumber\\
&\mapsto& \left(x,~x-y-a_5~;~a_6,a_7- 2a_5^2a_6,
~-a_3,a_4,a_5,a_1,a_2+2a_1a_5^2 \right).
\end{eqnarray*}

Similarly the action of $\sigma_{12}$
on $X$ and the space of parameters is
\begin{eqnarray*}
\sigma_{12}:&&(x,y;a_1,a_2,a_3,a_4,a_5,a_6,a_7)\nonumber\\
&\mapsto&\left(-y,~-x~;~-a_3,-a_4,
~-a_1,-a_2,a_5,-a_6,a_7-4a_5^2a_6 \right).
\end{eqnarray*}

The action of $w_1$ or $w_3$ is determined by the relation
$w_1=\sigma_{12}\circ w_2\circ  \sigma_{12}$ or 
$w_3=\sigma_{13}\circ w_1\circ  \sigma_{13}$ respectively as follows
\begin{eqnarray*}
w_1:&&(x,y;a_1,a_2,a_3,a_4,a_5,a_6,a_7)\nonumber\\
&\mapsto&\left(x-\frac{a_1}{y^2}-\frac{a_2}{a_1y},y~;~-a_1,a_2,
~a_3,a_4-\frac{2a_2a_3}{a_1},a_5,a_6,a_7+\frac{2a_2a_6}{a_1} \right).
\end{eqnarray*}
{\small
\begin{eqnarray*}
&&w_3:(x,y;a_1,a_2,a_3,a_4,a_5,a_6,a_7)\nonumber\\
&&\mapsto\left(
x-\frac{a_6}{(x-y-a_5)^2}-\frac{a_7-2a_5^2a_6}{a_6(x-y-a_5)},~
y-\frac{a_6}{(x-y-a_5)^2}-\frac{a_7-2a_5^2a_6}{a_6(x-y-a_5)}~;~\right. 
\nonumber\\ 
&&\left. a_1,a_2+ \frac{2 a_1(a_7-2a_5^2a_6)}{a_6},
a_3,a_4+\frac{2 a_3(a_7-2a_5^2a_6)}{a_6},a_5,-a_6,a_7-4a_5^2a_6 \right).
\end{eqnarray*}}

\subsection{The non-autonomous HV equation}

The composition $w_2 \circ \sigma_{13} \circ \sigma_{12}$ is reduced to
\begin{eqnarray}\label{nhv}
&&w_2 \circ \sigma_{13} \circ \sigma_{12}
:(x,y;a_1,a_2,a_3,a_4,a_5,a_6,a_7)\nonumber\\
&\mapsto&(-y,x-y-a_5-\frac{a_1}{y^2}-\frac{a_2}{a_1y}~;~ -a_6,
a_7-2a_5^2a_6-\frac{2a_2a_6}{a_1},\\ 
&&-a_1,-a_2,a_5,-a_3,
-a_4-2a_3a_5^2+\frac{2a_2a_3}{a_1})\nonumber
\end{eqnarray}
where $a_5$ can be normalized to be $a_5=0$ or $1$.

Of course this mapping satisfies the singularity confinement criterion
by construction and
in the case of $a_2=a_4=a_5=a_7=0$ and $a_1=a_3=a_6=a$ it 
coincides with the HV eq.(\ref{hv}) except their signs.
The difference between them comes from the assumption 
$\overline{a_5}=a_5$.
Assuming
$\overline{a_5}=-a_5$ under the actions of $w_2$, $\sigma_{13}$ and 
$\sigma_{12}$, 
we have $-w_2$, $-\sigma_{13}$ and $-\sigma_{12}$ as new 
$w_2$, $\sigma_{13}$ and $\sigma_{12}$
and therefore (\ref{nhv}) becomes as follows
\begin{eqnarray*}
&&w_2 \circ \sigma_{13} \circ \sigma_{12}
:(x,y;a_1,a_2,a_3,a_4,a_5,a_6,a_7)\nonumber\\
&\mapsto&(y,-x+y+a_5+\frac{a_1}{y^2}+\frac{a_2}{a_1y}~;~\\
&& a_6,-a_7+2a_5^2a_6+\frac{2a_2a_6}{a_1},a_1,a_2,-a_5,a_3,
a_4+2a_3a_5^2-\frac{2a_2a_3}{a_1}).\nonumber
\end{eqnarray*}
Actually in the case of $a_2=a_4=a_5=a_7=0$ and $a_1=a_3=a_6=a$ 
it coincides with the HV eq.(\ref{ihv}).

\section{Some other examples}

We present some examples of rational mappings which 
satisfy the singularity confinement criterion and some of which
have positive algebraic entropy. 

\subsection{Example 1}

Let $w_1,w_2,w_3$ and $a_i$ be as in \S~4.
First we consider the mapping $w_1\circ w_2$ 
{\small
\begin{eqnarray*}
&&w_1\circ w_2:(x,y;a_1,a_2,a_3,a_4,a_5,a_6,a_7)\\
&\mapsto&
\left(x - \frac{a_1}{y^2}- \frac{a_2}{a_1 y},
 y- \frac{a_3}{(x-a_1/y^2-a_2/(a_1 y))^2} -
\frac{a_4-2 a_2 a_3/a_1}{a_3
(x-a_1/y^2-a_2/(a_1 y))},\right.
~;~\\
&& \left. -a_1,a_2+ \frac{2(a_1a_4 -2 a_2a_3)}{a_3},-a_3,a_4-
 \frac{2 a_2 a_3}{a_1} ,a_5, a_6, 
a_7+\frac{2a_2a_6}{a_1}+ 2(a_4-2\frac{a_2a_3}{a_1})\frac{a_6}{a_3} \right)
\end{eqnarray*}
}
This mapping has the following properties. \\

\noindent 1)This mapping satisfies the singularity confinement criterion. \\
\noindent 2)The order of the $n$th iterate of mapping is $O(n^2)$ 
(easily seen from the action on the root lattice $<\alpha_i>$ 
\cite{takenawa2}).\\
\noindent 3)The actions on the parameters $a_5,a_6,a_7$ can be ignored. \\

This mapping is nothing but one of the discrete Painlev\'{e} equations,
since the surface obtained by blowing down the curves 
$E_9,E_{10},\cdots,E_{14}$ in $X$
is also the space of initial values and the type of Dynkin diagram 
corresponding to the irreducible components of 
anti-canonical divisor is $D_7^{(1)}$  
with the symmetry $A_1^{(1)}$.
Actually the irreducible components of 
anti-canonical divisor are
\begin{eqnarray*}
&&E_1-E_2,~E_2-E_3,~E_3-E_4,~E_5-E_6,~E_6-E_7,~E_7-E_8,\\
&&H_0-E_1-E_2,~H_1-E_5-E_6
\end{eqnarray*}
and the root basis of orthogonal lattice is
\begin{eqnarray*}
\alpha_1&=&2H_1-E_1-E_2-E_3-E_4,\\
\alpha_2&=&2H_0-E_5-E_6-E_7-E_8.
\end{eqnarray*}

Next we consider the mapping $w_3\circ w_2\circ w_1$.
This mapping is almost identical to the nonautonomous HV eq. after 
$3$ steps. Actually the latter becomes $w_2\circ w_3\circ w_1$.

At last we consider the mapping $w_2\circ w_3\circ w_2\circ w_1$.
This mapping satisfies the singularity confinement criterion and 
its algebraic entropy is $17+12\sqrt{2}$.

\subsection{Example 2}

We consider the following diagram as irreducible components
of the anti-canonical divisor.

\begin{picture}(400,100)
\put( 50,20){\line(1,0){20}}
\put( 75,20){\line(1,0){20}}
\put(100,20){\line(1,0){20}}
\put(125,20){\line(1,0){20}}
\put(150,20){\line(1,0){20}}
\put(175,20){\line(1,0){20}}
\put(200,20){\line(1,0){20}}
\put(225,20){\line(1,0){20}}
\put( 75,45){\line(1,0){20}}
\put(97.5,22.5){\line(0,1){20}}
\put(147.5,22.5){\line(0,1){20}}
\put(147.5,47.5){\line(0,1){20}}
\put(197.5,22.5){\line(0,1){20}}
\put(150,70){\line(1,0){20}}
\put(125,70){\line(1,0){20}}
\put(175,70){\line(1,0){20}}
\put(100,70){\line(1,0){20}}
\put(200,45){\line(1,0){20}}
\put(47.5,20){\circle{5}}
\put(72.5,20){\circle{5}}
\put(72.5,45){\circle{5}}
\put(97.5,70){\circle{5}}
\put(97.5,45){\circle{5}}
\put(97.5,20){\circle{5}}
\put(122.5,20){\circle*{5}}
\put(122.5,70){\circle{5}}
\put(147.5,20){\circle{5}}
\put(147.5,45){\circle*{5}}
\put(147.5,70){\circle{5}}
\put(172.5,20){\circle*{5}}
\put(172.5,70){\circle{5}}
\put(197.5,20){\circle{5}}
\put(197.5,45){\circle{5}}
\put(197.5,70){\circle{5}}
\put(222.5,20){\circle{5}}
\put(222.5,45){\circle{5}}
\put(247.5,20){\circle{5}}

\put(290,50){\circle{5}}
\put(290,35){\circle*{5}}
\put(320,50){\makebox(0,0){$-2$ curve}}
\put(320,35){\makebox(0,0){$-4$ curve}}
\put(275,20){\line(1,0){20}}
\put(330,20){\makebox(0,0){intersection}}
\end{picture}

This diagram is realized by the sequence of blow-ups from 
${\mathbb P}^1\times{\mathbb P}^1$ as follows
{\small
\begin{eqnarray*}
(x,y) &\maplleft{\mu_1}{(\infty,0)}& \left(\frac{1}{xy},y\right)
\maplleft{\mu_2}{(0,0)} \left(\frac{1}{xy^2},y\right)
\maplleft{\mu_3}{(0,0)} \left(\frac{1}{xy^2},xy^3\right)\\
&\maplleft{\mu_4}{(0,a_1)}& \left(\frac{1}{xy^2}, xy^2(xy^3-a_1)\right)
\maplleft{\mu_5}{(0,a_2)} 
\left(\frac{1}{xy^2},xy^2(xy^2(xy^3-a_1)-a_2)\right)\\
&\maplleft{\mu_6}{(0,a_3)}& 
\left(\frac{1}{xy^2},xy^2(xy^2(xy^2(xy^3-a_1)-a_2)-a_3)\right),\\
\end{eqnarray*}
\begin{eqnarray*}
(x,y) &\maplleft{\mu_7}{(0,\infty)}& \left(x,\frac{1}{xy}\right)
\maplleft{\mu_8}{(0,0)} \left(x,\frac{1}{x^2y}\right)
\maplleft{\mu_9}{(0,0)} \left(x^3y,\frac{1}{x^2y}\right)\\
&\maplleft{\mu_{10}}{(a_4,0)}& \left(x^2y(x^3y-a_4),\frac{1}{x^2y}\right)
\maplleft{\mu_{11}}{(a_5,0)} 
\left(x^2y(x^2y(x^3y-a_4)-a_5),\frac{1}{x^2y}\right)\\
&\maplleft{\mu_{12}}{(a_6,0)}& 
\left(x^2y(x^2y(x^2y(x^3y-a_4)-a_5),\frac{1}{x^2y}\right)
\end{eqnarray*}
}
and
{\small
\begin{eqnarray*}
(x,y) &\maplleft{\mu_{13}}{(\infty,\infty)}&
\left(\frac{1}{x},\frac{x}{y}\right)
\maplleft{\mu_{14}}{(0,1)}\left(\frac{1}{x},x(\frac{x}{y}-1)\right)
\maplleft{\mu_{15}}{(0,a_7)}
\left(\frac{1}{xz}, z\right),
\end{eqnarray*}
}
where we denote $z:=x(x/y-1)-a_7$,
{\small
\begin{eqnarray*}
&\maplleft{\mu_{16}}{(0,0)}&
\left(\frac{1}{xz^2}, z\right)
\maplleft{\mu_{17}}{(0,0)}
\left(\frac{1}{xz^2}, xz^3\right)
\maplleft{\mu_{18}}{(0,a_8)} 
\left(\frac{1}{xz^2},
 xz^2(xz^3-a_8)\right)\\
&\maplleft{\mu_{19}}{(0,a_9)}&
\left(\frac{1}{xz^2},xz^2(xz^2(xz^2-a_8)-a_9) \right)\\
&\maplleft{\mu_{20}}{(0,a_{10})}&
\left(\frac{1}{xz^2},xz^2(xz^2(xz^2(xz^3-a_8)-a_9)-a_{10}) \right).
\end{eqnarray*}
}
Similar to the case of the HV eq.(\S~4~), we obtain
{\small
\begin{eqnarray*}
w_2&:&(x,y: a_1,a_2,\cdots,a_{10})\\
&\mapsto& \left(x,y-\frac{a_4}{x^3}-\frac{a_5}{a_4x^2}-
(\frac{a_6}{a_4^2}-\frac{a_5^2}{a_4^3}) \frac{1}{x}~:~\right.\\
&&\left.a_1,a_2,a_3+\frac{3a_1^2a_5^2}{a_4^3}-\frac{3a_1^2a_6}{a_4^2},
-a_4,a_5,-a_6,a_7,a_8,a_9,
a_{10}-\frac{3a_5^2a_8^2}{a_4^3}+\frac{3a_6a_8^2}{a_4^2}
\right),
\end{eqnarray*}
\begin{eqnarray*}
\sigma_{13}&:&(x,y;a_1,a_2,\cdots,a_{10})\\
&\mapsto& \left(x,x-y-a_7~;~
a_8,a_9,a_{10}-3a_7^2a_8^2,-a_4,a_5,-a_6,a_7,a_1,a_2,a_3+3a_1^2a_7^2
\right),
\end{eqnarray*}
}
{\small
\begin{eqnarray*}
\sigma_{12}&:&(x,y;a_1,a_2,\cdots,a_{10})\\
&\mapsto&
\left(-y,-x~;~a_4,-a_5,a_6,a_1,-a_2,a_3,a_7,-a_8,a_9,-a_{10}+6a_7^2a_8^2
\right)
\end{eqnarray*}
}
and finally 
{\small
\begin{eqnarray*}
&&w_2\circ\sigma_{13}\circ\sigma_{12}:(x,y;a_1,a_2,\cdots,a_{10})~\mapsto\\
&&
\left(-y,- y+x-a_7-\frac{a_1}{y^3}-\frac{a_2}{a_1 y^2}+ 
  (\frac{-a_3}{a_1^2}+\frac{a_2^2}{a_1^3})\frac{1}{y}~;~~-a_8,a_9, \right. \\
&& \left.
-a_{10}+3a_8^2 a_7^2-\frac{3a_2^2a_8^2}{a_1^3}+\frac{3a_8^2a_3}{a_1^2},
a_1,-a_2,a_3,a_7,a_4,-a_5,
a_6+3a_4^2 a_7^2 +\frac{3a_2^2a_4^2}{a_1^3}-\frac{3a_3a_4^2}{a_1^2}
\right)
\end{eqnarray*}
}

In the case
$a_i=0$ for $ i=2,3,5,6,9,10$, this mapping reduces to 
\begin{eqnarray*}
&w_2\circ\sigma_{13}\circ\sigma_{12}:&(x,y;a_1,a_4,a_7,a_8)\\
&\mapsto&(-y,x-y-a_7+\frac{a_1}{y^3};-a_8,a_1,a_7,a_4).
\end{eqnarray*} 

We present some basic properties of this mapping.

The Picard group of the space of initial values is
$${\rm Pic}(X)= {\mathbb Z}H_0 + {\mathbb Z}H_1+  {\mathbb Z}E_1+\cdots+{\mathbb Z}E_{20}$$
and  the canonical divisor of $X$ is
$$K_X=-2H_0 -2H_1 + E_1+\cdots+E_{20}.$$
The irreducible components of the anti-canonical divisor are
\begin{eqnarray*}
&E_1-E_2,~E_2-E_3,~E_3-E_4,~E_4-E_5,~E_5-E_6,\\
&E_6-E_7,~E_7-E_8,~E_8-E_9,~E_9-E_{10},~E_{11}-E_{12},\\
&E_{12}-E_{13},~E_{15}-E_{16},~E_{16}-E_{17},\cdots,E_{19}-E_{20},\\
&H_0-E_1-E_2-E_3-E_{13},~H_1-E_5-E_6-E_7-E_{13},~
E_{14}-E_{15}-E_{16}-E_{17}
\end{eqnarray*}
and the root basis is
\begin{eqnarray*}
\alpha_1&=&3H_1-E_1-E_2-E_3-E_4-E_5-E_6,\\
\alpha_2&=&3H_0-E_7-E_8-E_9-E_{10}-E_{11}-E_{12},\\
\alpha_3&=&3H_0+3H_1-3E_{13}-3E_{14}-E_{15}-E_{16}-E_{17}-E_{18}-E_{19}-E_{20}.
\end{eqnarray*}
The Cartan matrix $2(\alpha_i\cdot\alpha_j)/(\alpha_i\cdot\alpha_i)$ is
\begin{eqnarray}\left[
\begin{array}{rrr}
2&-3&-3\\-3&2&-3\\-3&-3&2
\end{array}
\right].
\end{eqnarray}
This Cartan matrix is not finite, affine nor hyperbolic type.\\
The algebraic entropy is $2+\sqrt{3}$.\\

{\noindent{\bf Acknowledgment.}}
The author would like to thank J. Satsuma, H. Sakai, T. Tokihiro, K.Okamoto, 
R. Willox, A. Nobe, T. Tsuda and M. Eguchi 
for discussions and advice.

%%%%%%%%%%%%%%%%%%%%%%%%%%%

\appendix

\section{Uniqueness of the decomposition of the anti-canonical divisor}
\begin{theorem}
Let $X$ be the space of initial values of the HV eq. obtained in Section 2.
The anti-canonical class of divisors $-K_X$ can be reduced uniquely  
to prime divisors as
\begin{eqnarray*}
&&D_0+2D_1+D_2+D_3+ 2D_4+D_5\\
&&+3D_6+D_7+2D_8+D_9+2D_{10}+2D_{11}+2D_{12}
\end{eqnarray*}
\end{theorem}

\bp
Suppose that $-K_X$ is reduced to prime divisors as
$-K_X=\sum_{i=1}^m f_iF_i$, where $F_i$ is a prime divisor and 
$f_i$ is a positive integer. 
$F_i$ has the form
$$h_0^{(i)}H_0+h_1^{(i)}H_1-\sum_{j=1}^{14}e_j^{(i)}E_j,$$
where $h_j$ and $e_j$ are nonnegative integers and $h_j \leq2$
and moreover $h_0$ or $h_1$ is strictly positive
(remember that $-K_X=2H_0+2H_1-\sum_{j=1}^{14}E_j$), or 
in the case the curve is included in the total transform of a
blow-up point:
\begin{eqnarray*}
E_1-E_2, E_2-E_3, E_3-E_4, E_4, E_5-E_6, E_6-E_7, E_7-E_8, E_8,\\
E_9-E_{10},  E_{10}- E_{11}-E_{12}, E_{11}-E_{12},
E_{12}-E_{13}, E_{13}-E_{14},  E_{14}. 
\end{eqnarray*}

Let us first suppose that there does not exist $F_i$ such 
that $F_i=E_{13}-E_{14}$.
Then there exists a $F_i$ such that $F_i$ has the form as 
$h_0H_0+h_1H_1-\sum_{j=1}^{14}e_jE_j$ where $e_{14}$ is strictly 
positive. This 
means that $F_i$ has an intersection point with $E_{14}$ and therefore
this divisor passes the point of the $14$-th blow-up before the blow-up. 
Namely, 
$$\left(\frac{1}{x^2(x/y-1)},
 x^2(\frac{x}{y}-1)(x^3(\frac{x}{y}-1)^2-a)\right)=(0,0)$$
as $x,y\to \infty$. Denoting $u=1/x,v=1/v$, we have
$u^3/(v-u)=0$ and $(v-u)(v^2-2v+1-a u^5)/u^8=0$ as $u,v\to 0$.
This implies $v^2-2v+1-au^5=o(\frac{u^8}{v-u})$ and hence we have
$h_0\geq 5$, which is a contradiction. 

Similarly there does not exist a divisor which has the form
$h_0H_0+h_1H_1-\sum_{j=1}^{13}e_jE_j$ where $e_{12}$ or $e_{13}$ is 
strictly positive and $h_j\leq 2$.\\

Now, we can consider the case where there does not exist a $F_i$ such 
that $F_i=E_{14}$.
Then according to the above result, 
there should exist integers $j,k$ such that
$F_j=E_{13}-E_{14}, F_k=E_{12}-E_{13}$ and
$f_j= 1, f_k= 2$ (in the absence of a prime divisor $E_{14}$,
the form of $-K_X$ forces $f_j=1$ and subsequently $f_k=2$
in order to have the correct $E_{13}$ dependence).
Considering the new coefficient of $E_{12}$, 
we have that the sum of the coefficients of $E_{11}-E_{12}$ and 
$E_{10}-E_{11}-E_{12}$ has to be $3$.

Then we have the following possibilities: \\
i) the coefficient of $E_{11}-E_{12}$ is $0$;\\
ii) the coefficient of $E_{11}-E_{12}$ is $1$; \\
iii) the coefficient of $E_{11}-E_{12}$ is $2$; \\
iv) the coefficient of $E_{11}-E_{12}$ is $3$. 

In the case i),ii),iii) or iv) the coefficient of $E_{11}$ is 
$-3,-1,1$ or $3$ respectively. Hence i) is impossible. 
To pass the point of the $11$-th blow-up, a divisor whose class has the form
$h_0H_0+h_1H_1-\sum_{j=1}^{11}e_jE_j$ must satisfy
the equation 
$u=0$ and $(v-u)/u^2=0$. This implies $h_0\geq 2, h_1\geq 1$ and $e_{11}=1$
and therefore iii) and iv) are impossible.    
Consequently the coefficient of $E_{11}-E_{12}$ has to be $1$ 
and the coefficient of $E_{10}-E_{11}-E_{12}$  has to be $2$.

Along the same lines, considering the coefficient of $E_{10}$,
we have the following possibilities: \\
\noindent a) the coefficient of $E_9-E_{10}$ is $0$, in which case 
the coefficient of $E_9,E_{10}$ is $0,2$;\\
\noindent b) the coefficient of $E_9-E_{10}$ is $1$, in which case 
the coefficient of $E_9,E_{10}$ is $1,1$; \\
\noindent c) the coefficient of $E_9-E_{10}$ is $2$, in which case 
the coefficient of $E_9,E_{10}$ is $2,0$; \\
\noindent d) the coefficient of $E_9-E_{10}$ is $3$, in which case 
the coefficient of $E_9,E_{10}$ is $3,-1$.
  
To pass the point of the $10$-th blow-up, a divisor whose class has the form
$h_0H_0+h_1H_1-\sum_{j=1}^{11}e_jE_j$ must satisfy
the equation 
$u=0$ and $v/u=1$. It implies two possibilities:
i) $h_0\geq 1, h_1\geq 1$, $e_9=e_{10}=1$
and $e_2,e_3,e_4,e_6,e_7,e_8=0$
(since if $e_2$ would be strictly positive, the intersection of the divisor
$h_0H_0+h_1H_1-\sum_{j=1}^{11}e_jE_j$ and $x=\infty$ ($=H_0$) is greater 
than or equal to $3$, which is impossible)   
or ii) $h_0H_0+h_1H_1-\sum_{j=1}^{11}e_jE_j=2H_0+2H_1-2E_9-2E_{10}$.
Therefore a) is impossible. In the case b) the coefficient
of $E_9$ becomes $-1$ (or $1,3$ in the cases c), d) respectively)  
and the coefficients of $H_0$ and $H_1$ become
equal to $2$ (or become greater than or equal to $1$, 
become $0$ in the cases c), d) respectively).

In the case b) the coefficients of $E_1,E_2,\cdots,E_8$  
is $0$ and hence the remaining contribution 
$-E_1-\cdots-E_{8}$ must be a sum of the terms: 
\begin{eqnarray}\label{appa}
E_1-E_2, E_2-E_3, E_3-E_4, E_4, E_5-E_6, E_6-E_7, E_7-E_8, E_8
\end{eqnarray}
with nonnegative integer coefficients. Obviously this is impossible.
In the case c) the remaining contribution has the form
$b_0'H_0+b_1'H_1-b_1E_1-\cdots-b_8E_8-2E_9$, where 
$b_0',b_1'=0 \mbox{ or } 1$, $b_2,b_3,b_4,b_6,b_7,b_8=1$
and $b_1,b_2= 0 \mbox{ or } 1$ and
hence it must be a sum of $H_0-E_1-E_2-E_9$, 
$H_1-E_5-E_6-E_9$ and (\ref{appa}) with nonnegative integer coefficients. 
By straight forward discussion it can be seen that this case is 
also impossible.

The coefficient of $E_9-E_{10}$ therefore has to be equal to $3$
and hence the coefficient of $E_9$ has to be $3$ as well,
from which point on it can be easily seen that the remaining 
contributions to $-K_X$ are uniquely determined.

Similarly, starting from the supposition that
there does exist a $F_i$ such 
that $F_i=E_{14}$, similar proof can be given. \ep

%%%%%%%%%%%%%%%%%%%%


\begin{thebibliography}{9}

\bibitem{arnord}
Arnold, V.I.: {Dynamics of complexity of intersections}.
Bol. Soc. Bras. Mat. \textbf{21}, 1--10 (1990)

\bibitem{bv}
Bellon, M.P. and Viallet, C.M. {Algebraic entropy}.
Commun. Math. Phys. \textbf{204}, 425--437 (1999)

\bibitem{cd}
Cossec, F., Dolgachev, I.: {\it Enriques surfaces I}.
Boston: Birkh\"{a}user 1988

\bibitem{dolgachev}
Dolgachev, I.:{Weyl groups and Cremona transformations}.
Proc. Symp. Pure Math. \textbf{40}, 283--294 (1983)

\bibitem{dolgachev2}
Dolgachev, I.:{\it Point sets in projective spaces and theta functions}.
Ast\'{e}risque  Soc.Math.de France \textbf{165}, 1988


\bibitem{grp}
Grammaticos, B., Ramani, A. and Papageorgiou, V.:
{Do integrable mappings
  have the Painlev\'{e} property?}
Phys.\ Rev.\ Lett. \textbf{67}, 1825--1827  (1991)

\bibitem{hartshorne}
Hartshorne, R. 
{\it Algebraic geometry}.
New York: Springer-Verlag 1977

\bibitem{hv}
Hietarinta, J. and Viallet, C.M.:
{Singularity confinement and chaos in discrete systems}.
Phys. Rev. Lett. \textbf{81}, 325-328 1997

\bibitem{js}
Jimbo, M. and Sakai, H.:
{A {\it q}-analog of the sixth Painlev\'{e}
equation}.
Lett. Math. Phys. \textbf{38}, 145--154 (1996)

\bibitem{kac}
Kac, V.:
{Infinite dimensional lie algebras, 3rd ed.}.
Cambridge: Cambridge University Press 1990

\bibitem{looijenga}
Looijenga, E.:
{Rational surfaces with an anti-canonical cycle}.
Annals of Math. \textbf{114}, 267--322 (1981)


\bibitem{otgr}
Ohta, Y., Tamizhmani, K.M., Grammaticos, B. and Ramani, A.:
{Singularity confinement and algebraic entropy: the case of the discrete
Painlev\'e  equations}.
Phys. Lett. A \textbf{262}, 152--157  (1999)

\bibitem{okamoto}
Okamoto, K.:
{Sur les feuilletages associ\'{e}s aux \'{e}quations du second 
ordre \`{a} points critiques fixes de P.Painlev\'{e}. (French)}.
Japan J. Math. \textbf{5}, 1--79  (1979)


\bibitem{rgh}
Ramani, A., Grammaticos, B. and Hietarinta, J.: 
{Discrete versions of the
  Painlev\'{e} equations}.
Phys. Rev. Lett. \textbf{67}, 1829--1832 (1991)


\bibitem{sakai} Sakai, H.:
{Rational surfaces associated with affine root systems and geometry of
the Painlev\'{e} equations}.
Comm. Math. Phys. at press webpage:
 http://www.kusm.kyoto-u.ac.jp/preprint/preprint99.html


\bibitem{takenawa}
Takenawa, T.:
{A geometric approach to singularity confinement
and algebraic entropy}.
J. Phys. A: Math. Gen. \textbf{34} L95--L102 (2001)

\bibitem{takenawa2} 
Takenawa, T.:
{Algebraic entropy and the space of initial values for discrete 
dynamical systems}. nlin.SI/0103011



\bibitem{wan}
Wan, Z.: 
{\it Kac-Moody algebra}. Singapore: World Scientific 1991





\end{thebibliography}
\end{document}